\documentstyle[12pt,psfig]{article}
\makeatletter
\def\@cite#1#2{{[{#1}]\if@tempswa\typeout
{IJCGA warning: optional citation argument
ignored: `#2'} \fi}}

\newcount\@tempcntc
\def\@citex[#1]#2{\if@filesw\immediate\write\@auxout{\string\citation{#2}}\fi
 \@tempcnta\z@\@tempcntb\m@ne\def\@citea{}\@cite{\@for\@citeb:=#2\do
   {\@ifundefined
    {b@\@citeb}{\@citeo\@tempcntb\m@ne\@citea\def\@citea{,}{\bf ?}\@warning
     {Citation `\@citeb' on page \thepage \space undefined}}%
    {\setbox\z@\hbox{\global\@tempcntc0\csname b@\@citeb\endcsname\relax}%
    \ifnum\@tempcntc=\z@ \@citeo\@tempcntb\m@ne
   \@citea\def\@citea{,}\hbox{\csname b@\@citeb\endcsname}%
     \else
      \advance\@tempcntb\@ne
      \ifnum\@tempcntb=\@tempcntc
      \else\advance\@tempcntb\m@ne\@citeo
      \@tempcnta\@tempcntc\@tempcntb\@tempcntc\fi\fi}}\@citeo}{#1}}
\def\@citeo{\ifnum\@tempcnta>\@tempcntb\else\@citea\def\@citea{,}%
  \ifnum\@tempcnta=\@tempcntb\the\@tempcnta\else
   {\advance\@tempcnta\@ne\ifnum\@tempcnta=\@tempcntb \else
\def\@citea{--}\fi
   \advance\@tempcnta\m@ne\the\@tempcnta\@citea\the\@tempcntb}\fi\fi}
\makeatother

\def\NPB#1#2#3{{\it Nucl.~Phys.} {\bf{B#1}} (19#2) #3}

\newcommand{\mh}{M_{h^0}}
\newcommand{\Hh}{\lower1.2ex\hbox{$\stackrel{\textstyle
H}{\footnotesize\sim}$}}
\newcommand{\Hho}{\lower1.2ex\hbox{$\stackrel{\textstyle
H_1}{\footnotesize\sim}$}}
\newcommand{\Hhw}{\lower1.2ex\hbox{$\stackrel{\textstyle
H_2}{\footnotesize\sim}$}}
\newcommand{\h}{\lower1.2ex\hbox{$\stackrel{\textstyle
h}{\footnotesize\sim}$}}
\newcommand{\dr}{\mbox{\footnotesize$\overline{\rm DR}$~}}
\newcommand{\ms}{\mbox{\footnotesize$\overline{\rm MS}$~}}
\newcommand{\olf}{16\pi^2}

\newcommand{\hxt}{{\hat{X}}_t}

\newcommand{\hyt}{{\hat{Y}}_t}

\newcommand{\tilt}{\tilde{t}}
\newcommand{\sto}{\tilde{t}_1}
\newcommand{\stw}{\tilde{t}_2}
\newcommand{\str}{\tilde{t}_R}

\newcommand{\lt}{\ln{m_t^2\over Q^2}}
\newcommand{\ls}{\ln{M^2_S\over Q^2}}

\newcommand{\nn}{\nonumber}

\newcommand{\bmt}{\overline{m}_t}

\newcommand{\gsim}{\lower.7ex\hbox{$\;\stackrel{\textstyle>}{\sim}\;$}}
\newcommand{\lsim}{\lower.7ex\hbox{$\;\stackrel{\textstyle<}{\sim}\;$}}
\newcommand{\be}{\begin{equation}}
\newcommand{\ee}{\end{equation}}

\def\baselinestretch{1}
\begin{document}
\catcode`@=11
\newtoks\@stequation
\def\subequations{\refstepcounter{equation}%
\edef\@savedequation{\the\c@equation}%
  \@stequation=\expandafter{\theequation}
  \edef\@savedtheequation{\the\@stequation}
  \edef\oldtheequation{\theequation}%
  \setcounter{equation}{0}%
  \def\theequation{\oldtheequation\alph{equation}}}
\def\endsubequations{\setcounter{equation}{\@savedequation}%
  \@stequation=\expandafter{\@savedtheequation}%
  \edef\theequation{\the\@stequation}\global\@ignoretrue

\noindent}
\catcode`@=12
\begin{titlepage}

\title{{\bf RADIATIVE CORRECTIONS\\
TO THE HIGGS BOSON MASS FOR A\\
 HIERARCHICAL STOP SPECTRUM}}
\vskip2in
\author{  
{\bf J.R. Espinosa$^{1,2}$\footnote{\baselineskip=16pt E-mail: {\tt
espinosa@makoki.iem.csic.es}}} and 
{\bf I. Navarro$^{3}$\footnote{\baselineskip=16pt E-mail: {\tt
ignacio@makoki.iem.csic.es}}}
\hspace{3cm}\\
 $^{1}$~{\small I.M.A.F.F. (CSIC), Serrano 113 bis, 28006 Madrid, Spain}
\hspace{0.3cm}\\
 $^{2}$~{\small I.F.T. C-XVI, U.A.M., 28049 Madrid, Spain}
\hspace{0.3cm}\\
 $^{3}$~{\small I.E.M. (CSIC), Serrano 123, 28006 Madrid, Spain}.
} 
\date{} 
\maketitle 
\def\baselinestretch{1.15} 
\begin{abstract}
\noindent 
An effective theory approach is used to compute analytically the radiative
corrections to the mass of the light Higgs boson of the Minimal
Supersymmetric Standard Model when there is a hierarchy in the masses of
the stops ($M_{\sto}\gg M_{\stw} \gg M_{\mathrm top}$, with moderate stop
mixing).  The calculation includes up to two-loop leading and
next-to-leading logarithmic corrections dependent on the QCD and top-Yukawa
couplings, and is further completed by two-loop non-logarithmic corrections
extracted from the effective potential. The results presented disagree 
already at two-loop-leading-log level with
widely used findings of previous literature. Our formulas can be
used as the starting point for a full numerical resummation of logarithmic
corrections to all loops, which would be mandatory if the hierarchy between
the stop masses is large.
\end{abstract}

\thispagestyle{empty}
\vspace*{3cm}
\leftline{March 2001}
\leftline{}

\vskip-22cm
\rightline{}
\rightline{IFT-UAM/CSIC-01-11}
\rightline{IEM-FT-214/01}
\rightline{hep-ph/0104047}
\vskip3in

\end{titlepage}
\setcounter{footnote}{0} \setcounter{page}{1}
\newpage
\baselineskip=20pt

\noindent

\section{Introduction}

The Minimal Supersymmetric Standard Model (MSSM) predicts a light Higgs boson,
with mass $M_{h^0}$ of the order of the scale of electroweak symmetry breaking
($G_F^{-1}\sim v=246$ GeV) times a small Higgs quartic-self-coupling. That
Supersymmetry (SUSY) can naturally trigger this breaking, and stabilize the
scale at which it takes place, is the most interesting part of the story (see
\cite{SUSY} for reviews and references). Here we take that for granted and our
focus is on the perturbatively small coupling. Its smallness comes about
because of two reasons: first, Supersymmetry dictates that the Higgs quartic
self-couplings are given by gauge couplings (from $D$-terms) and by
superpotential Yukawa couplings (from $F$-terms); second, the latter $F$-term
contributions are absent in the MSSM since, in this model, quantum numbers
prevent superpotential terms cubic in the Higgs fields. It is generic
\cite{bound} that quartic Higgs couplings are directly related to the Higgs
mass after electroweak symmetry breaking (the Standard Model is the best known
example). All this results in the well known tree-level upper bound
$M_{h^0}^2\leq M_Z^2\cos^22\beta$ [the SUSY parameter $\tan\beta$ is the ratio
$v_2/v_1$ of the two Higgs vacuum expectation values (vevs). We follow the
usual convention: $v_2$ generates the mass of the top quark and $v_1$ that of
the bottom quark].

Radiative corrections to $M_{h^0}$ can be quite important because of
top-stop loops that introduce a dependence on the top Yukawa coupling,
$h_t$, which is sizeable, while this coupling does not enter in the
tree-level Higgs mass. This can lead to cases in which one-loop radiative
corrections to $M_{h^0}$ are comparable to, or even larger than the
tree-level part of it (without this being an indication of the failure of
the perturbative expansion).

In addition, the one-loop corrections to $\mh$ are logarithmically
sensitive to the mass ratio, $m_{\tilde{t}}/m_t$, of the average stop mass
over the top mass, which could be large if there is a hierarchy,
$M_{SUSY}/M_{EW}\gg 1$, of the SUSY mass scale over the electroweak scale.  
As a consequence, radiative corrections to $\mh$ beyond one-loop can be
important if $\ln(m_{\tilde{t}}/m_t)$ is large. In that event, standard
renormalization group (RG) techniques can be used with advantage to resum
these logarithmic corrections to all loops.

During the last decade, the precise determination of $\mh$ as a function of
the supersymmetric parameters has received continued attention
\cite{higgs1}-\cite{CH3W2}. The development of increasingly refined
calculations of $\mh$ is the story of a stepwise climbing of this ladder of
loop corrections and has been told elsewhere (see {\it e.g.}
\cite{EZ2} for a brief account) so it will not be repeated here. The
current status of what has been achieved, by the combined use of direct
diagrammatic calculations, effective potential methods and RG techniques,
could be summarized in this way: all one-loop corrections are known
\cite{one,PBMZ} and the dominant two-loop corrections of order ${\cal
O}(\alpha_s\alpha_t m_t^2)$ and ${\cal O}(\alpha_t^2 m_t^2)$ are also
known, including finite (non-logarithmic) contributions [here
$\alpha_s\equiv g_s^2/(4\pi)$ and $\alpha_t\equiv h_t^2/(4\pi)$, with $g_s$
the QCD gauge coupling]. Higher order corrections at leading-log and
next-to-leading-log order [${\cal O}(\alpha_t
m_t^2\alpha_{t,s}^n[\ln(m_{\tilde{t}}/m_t)]^n)$ and ${\cal O}(\alpha_t
m_t^2\alpha_{t,s}^{n+1}[\ln(m_{\tilde{t}}/m_t)]^n)$] can be resummed using
one-loop and two-loop RG $\beta$-functions, respectively.
 
Beyond tree level, $\mh$ is sensitive to many SUSY parameters, but the most
important are those of the stop sector (and of the sbottom sector also for
large $\tan\beta$). They are given by the stop mass matrix:
\begin{equation}
{\bf M}^2_{\tilt}\
 \simeq\
\left[\begin{array}{cc}
M_L^2+m_t^2  &  m_tX_t\\[2mm]
m_t X_t^*      &  M_R^2+m_t^2
\end{array}\right]\ ,
\label{stopmat0}
\end{equation}  
where we have neglected $D$-terms, $M_L^2$ ($M_R^2$) is the soft-mass for
$\tilde{t}_L$ ($\tilde{t}_R$), and 
\be
X_t\equiv A_t+\mu^*/\tan\beta\ ,
\ee
with $A_t$ the soft trilinear coupling associated to the top Yukawa coupling
and $\mu$ the supersymmetric Higgs mass in the superpotential.

The dependence of the radiative corrections to $\mh$ on these parameters
has been studied before in different specific regimes. In this paper, we
focus on the case in which there is a double hierarchy, $M_L\gg M_R\gg m_t$
(the case $M_L\gg M_R\simeq m_t$ can be worked out along similar lines). In
this situation one should care, not only about potentially large logarithms
like the usual $\ln(M_L/m_t)$ and $\ln(M_R/m_t)$, but also about
$\ln(M_L/M_R)$. Radiative corrections to $\mh$ for this type of stop
spectrum have been considered in the past \cite{CQW,H3} but there is room
for improvement, as we will show. First, if the hierarchy between the stop
masses is large, a numerical resummation of logarithmic corrections to all
loops is necessary to get an accurate determination of the Higgs mass and,
in order to do this, one has to identify first the relevant RG functions and
threshold corrections. So far, this has not been done. Second, although 
previous analyses  represent important steps ahead, they are not complete
in one sense or another: either they do not include all potentially
relevant corrections or, if they do, the corrections are not cast in a 
form suitable for RG resummation.

The plan of the paper is the following. In section~2 we present the main
calculation. We use an effective theory method to extract and classify all
two-loop dominant ({\it i.e.} $h_t$ and $g_s$-dependent) radiative
corrections to $\mh$. The result of this calculation can be used as the 
starting point for a
full numerical evaluation of the Higgs mass in the case of hierarchical stop
spectra,
although we do not undertake that task in this paper. In section~3, we
discuss the possibility of finding a one-loop `improved' approximation to
$\mh$ that, playing with a judicious choice of the scales at which
parameters are evaluated, tries to absorb higher order corrections. This
exercise is a good point at which to compare  our main result,
presented in section~2, to previous analyses existing in the literature, with some
of which we disagree already at the level of two-loop leading-log
corrections. We dedicate section~4 to such comparisons. Section~5 presents
our conclusions and outlook for future work. For reference, Appendix~A
presents an explicit formula for $\mh$ which includes up to 
two-loop-next-to-leading logarithmic corrections.
Appendix B is devoted to the calculation of two-loop threshold corrections
for the Higgs quartic self-coupling, of direct interest for the
completeness of the two-loop calculation of $\mh$. Finally, Appendix~C
gives the relationships between $\ms$ running parameters (in which our
results are expressed) and on-shell (OS) quantities.

\section{Effective theory calculation}

We consider the MSSM with  a particle spectrum in which all supersymmetric
particles have a common mass, $M_{SUSY}$, much larger than the electroweak
scale (say a few TeV) except for the lightest stop, which is much lighter
although still heavier than the top quark. In particular, we remark that
the mass of the
pseudoscalar Higss, $M_{A^0}$, is also taken to be $M_{SUSY}$, and
therefore, the model contains just one light Higgs doublet. To be
precise, and referring to the
stop mass matrix written in eq.~(\ref{stopmat0}), we consider
\be
m_t\ll M_R\ll M_L=M_{SUSY}\ .
\label{hier}
\ee
Concerning stop mixing, we also assume that it is not too large, so that it
is a good approximation to say that the lightest stop is mostly\footnote{
This avoids problems with a large contribution to $\Delta\rho$, which would
be present in the opposite limit in which the light stop is mostly
$\tilde{t}_L$.} $\tilde{t}_R$, while the heavier one is mainly
$\tilde{t}_L$.  In other words, we are in a situation in which the stop
mixing angle is small. Nevertheless we do keep the dependence with the stop
mixing parameter $X_t$ and we will derive our results as a series in powers
of $m_tX_t/M_L^2$ (note that our approximation is $m_t X_t/M_L^2\ll 1$, not
$X_t/M_L\ll 1$). The case of a hierarchy in stop masses due to  very large
$X_t$ (rather than to different diagonal soft masses) is worth separate
study but it is more complicated and we do not consider it here.

To compute the radiatively corrected Higgs mass in the hierarchical case
(\ref{hier}), we make use of an effective lagrangian approach, descending
in energy from $M_{SUSY}$ down to the electroweak scale $m_t$. In doing so
we encounter different effective theories at different energy scales. Above
$M_L=M_{SUSY}$ the relevant theory is the full MSSM. Between $M_L=M_{SUSY}$
and $M_R$ the effective theory contains only the Standard Model particles
with a single Higgs doublet (that particular rotation of
the two Higgs
doublets of the MSSM which is responsible for electroweak symmetry breaking
and has SM properties) and, in addition, the light stop. Below the mass
scale $M_R$ of that light stop the effective theory is simply the pure
Standard Model (with calculable non-renormalizable operators, remnant of
the decoupling of heavy SUSY particles).

To compute $\mh$ we start at $M_{SUSY}$ with the known value of the quartic
Higgs coupling, $\lambda_H$, as a boundary condition fixed by Supersymmetry.
We run this coupling down to $m_t$ in the different effective theories just
mentioned, taking care of threshold corrections whenever some energy threshold
is crossed. The procedure is standard and follows the general prescriptions
for effective theory calculations. For general reviews of this subject
we refer to \cite{Efft} and references therein. We also found useful some
general discussions in
ref.~\cite{nyffeler}, a more specialized paper which studies the effective
theory of a linear $O(N)$ sigma model. Similar effective theory techniques
have been applied to study the decoupling limit of the MSSM with heavy
superpartners \cite{MSSMdec}.

We work in an approximation that neglects in radiative corrections all
couplings except $g_s$ and $h_t$ [our results could be extended easily to
include also $h_b$ (bottom-Yukawa) corrections, which can be significant
for large values of $\tan\beta$]\footnote{Following this approximation, we
neglect the effects of gauge couplings in the masses of SUSY particles
(which only affect $\mh$ through radiative corrections). In particular,
Higgsinos simply have mass $|\mu|=M_{SUSY}$.}. We keep electroweak gauge
couplings only in the tree-level contribution\footnote{For
numerical applications the full dependence on gauge couplings at one-loop is
known and can be included.} to $\mh$. In this connection,
the quartic Higgs coupling, $\lambda_H$, is considered to be itself of
one-loop order [$\lambda_H\sim h_t^4/(16\pi^2)$] when it appears in
radiative corrections. Within this approximation, we plan to extract
analytically the radiative corrections to $\mh$ up to two-loop order, that
is, we compute one-loop leading-log and finite terms plus two-loop
leading-log, next-to-leading-log and finite corrections to $\mh$. We also
use, whenever necessary, expansions in powers of the mass ratios $m_t/M_R$,
$m_t/M_L$ and $M_R/M_L$, as is common use in effective theory calculations.
If these ratios are not small there is no necessity of using
effective theory methods: the corresponding corrections to $\mh$ are not
logarithmically enhanced and other existing calculations should be valid.

The analytical result for $\mh$ that we obtain in this way is interesting
for two reasons: first, it identifies the ingredients (threshold
corrections and renormalization group functions)
necessary for a full numerical computation of the running of $\lambda_H$;
second, it is interesting in order to develop simple and compact
approximations to the
full numerical results.  Our goal is then to
find a two-loop formula for the Higgs mass 
in which the corrections are classified in such a way as
to permit a numerical resummation of leading and next-to-leading logarithmic
corrections to all loops. As explained, this resummation is mandatory if the
hierarchy (\ref{hier}) is sizeable, when simple analytical approximations
start
to fail. We defer that numerical evaluation of $\mh$ to a future
publication and concentrate here upon the analytical study.

\subsection{Plan}

The method we follow is very similar to that used by Haber and
Hempfling
in ref.~\cite{H2} to compute radiative corrections to $\mh$ for low values
of the pseudoscalar mass, $M_{A^0}$, case in which the theory below
$M_{SUSY}$ is a
two-Higgs-doublet model. The plan of our calculation is to integrate the
equation $d\lambda_H/d\ln Q^2=\beta_{\lambda_H}$ from $M_{SUSY}=M_L$,
where $\lambda_H$ is related to SUSY parameters, to $M_{EW}=m_t$, where
$\lambda_H$ determines the Higgs mass. Taking into account the different
running in the two effective theories, above and below the intermediate
threshold at $M_R$, and writing explicitly the threshold corrections to
$\lambda_H$ [with a 0 superindex, as in $\lambda_H^0(M_L)$, we always indicate
a tree-level value], we
find
\begin{eqnarray}
\lambda(m_t)&=&\lambda_H^0(M_L)+\delta\lambda_H(M_L)
-\int_{Q=m_t}^{M_R^-}\beta_\lambda(Q)d\ln
Q^2\nn\\
&&+\delta\lambda_H(M_R)
-\int_{Q=M_R^+}^{M_L}\beta_{\lambda_H}(Q)d\ln Q^2\ .
\end{eqnarray}
The quantities $\delta\lambda_H$ are the threshold corrections for
$\lambda_H$ at the indicated scales. We call $\lambda$ the Higgs quartic
coupling below $M_R$ to distinguish it from $\lambda_H$ above $M_R$. 

If we next expand the $\beta$-functions around a particular value of the scale,
and make a loop expansion up to two-loops [$\beta=\beta^{(1)}+\beta^{(2)}+...$]
we get 
\begin{eqnarray}
\lambda(m_t)&=&\lambda_H(M_L)+\delta\lambda_H(M_L)+\delta\lambda_H(M_R)\nn\\
&&-\left[\beta_\lambda^{(1)}(m_t)+\beta_\lambda^{(2)}\right]\ln{M_R^2\over
m_t^2}-\left[\beta_{\lambda_H}^{(1)}(M_R)+\beta_{\lambda_H}^{(2)}\right]
\ln{M_L^2\over M_R^2}\nn\\
&&-{1\over 2}{d\beta_\lambda^{(1)}\over d\ln Q^2}\ln^2{M_R^2\over m_t^2}
-{1\over 2}{d\beta_{\lambda_H}^{(1)}\over d\ln Q^2}\ln^2{M_L^2\over
M_R^2} 
+...  
\label{plan}
\end{eqnarray}
It is important to make explicit the scale at which one-loop
$\beta$-functions are evaluated, because different scale choices amount to
a two-loop difference [the scale choice in two-loop terms like
$\beta^{(2)}$ or $d\beta^{(1)}/d\ln Q^2$ has only effects starting at
three loops].  The same comments apply to
the choice of the scale at which to evaluate the masses inside one-loop
logarithms. In (\ref{plan}), the RG procedure dictates that they are
evaluated at a scale equal to the mass itself, that is, $M_R\equiv
M_R(M_R)$, $M_L\equiv M_L(M_L)$ and $m_t\equiv \overline{m}_t(m_t)$.
In a similar way, it is important that the couplings which
appear in $\beta^{(1)}$'s are evaluated taking into account the
corresponding one-loop threshold corrections.  All
this will be shown more explicitly in the following subsections.

Eq.~(\ref{plan}) already illustrates some properties of radiative
corrections which are generic: {\it i)} Leading-log contributions
at any order depend only on one-loop RG-functions and are, therefore,
insensitive to threshold corrections and two-loop or higher RG-functions.
The reason is simple: by definition, in leading-log corrections each power of
the loop
expansion parameter $\alpha$ ($\alpha_t$ or $\alpha_s$ in our case) is
 accompanied by a logarithm (that arises from RG running between two
mass scales). However, threshold corrections introduce powers of
$\alpha$'s without such logarithms, while $n^{th}$-order RG-functions
introduce a factor $\alpha^{n-1}$ for each $\alpha\log$.
{\it ii)} Next-to-leading-log terms are instead sensitive first to two-loop
RG-functions and second, to one-loop RG-functions times one-loop threshold
corrections. In turn, they are not sensitive to three-loop (or higher)
RG-functions or two-loop (or higher) threshold corrections. 

With this
hierarchical classification of radiative corrections in mind, our
calculation aims at finding the relevant one and two-loop RG-functions
plus one-loop threshold corrections. This would allow the resummation of
leading and next-to leading logarithmic contributions to $\mh$ to all
loops. Nevertheless, in our analytical formulas we stop at two-loops,
including also two-loop non-logarithmic terms.

\subsection{SUSY threshold: matching MSSM with SM
$+$ $\tilde{t}_R$}

The effective theory below the supersymmetric threshold at $M_{SUSY}$ is
described by the most general Lagrangian built with SM particles plus
$\tilde{t}_R$, non-renormalizable in general but invariant under
the $SU(3)_C\times SU(2)_L\times U(1)_Y$ gauge symmetry:
\begin{eqnarray}
{\cal L}_{SM+\str}&=&({\cal D}_\mu \str)^{\alpha\dagger}({\cal D}^\mu
\str)^\alpha+({\cal D}_\mu \h)^{\dagger}({\cal D}^\mu
\h)-\overline{M}_R^2|\str|^2-\overline{m}^2|\h|^2\nn\\
&&+[g_tQ_L^\alpha c\, t_R^{c\alpha}\cdot \h+{\mathrm h.c.}]-{1\over
2}\lambda_U |\str|^4-{1\over 2}\lambda_H
|\h|^4-\lambda_{HU} |\str|^2|\h|^2\nn\\
&&-\lambda_{HU}'(\str^{\alpha *}\h)\cdot
\partial^2 (\overline{\h}\str^\alpha)
-\kappa^2|\str|^2|\h|^4
-\chi |\h|^2|\str|^4 +... \ ,
\label{Lsmstr}
\end{eqnarray}
where the ellipsis stands for terms of higher order both in fields (bosonic
or fermionic) and derivatives. The Higgs doublet field is represented by
$\h$, $Q_L^\alpha$ is the top-bottom quark doublet and $t_R^{\alpha}$ the
right-handed top quark field ($\alpha$ is a colour index). The dot
($\cdot$) stands for the $SU(2)$
invariant product and $c=-i\sigma^2$. We have written explictly only the
third generation Yukawa coupling, which in this intermediate-energy theory we
call $g_t$. We keep only terms directly related to our calculation and do not
write, for example, fermion kinetic terms.

The parameters in the Lagrangian (\ref{Lsmstr}) are determined by matching at
the scale $Q=M_{SUSY}$ with the full MSSM theory, {\it i.e.} by requiring
that the effective theory and the full MSSM give the same physics at low
momentum \cite{Efft}. To do this matching at tree level, we first obtain the
equations of
motion of the heavy MSSM fields and substitute them (making a low momentum
expansion) in the MSSM Lagrangian. What one obtains is (we use a prime to
distinguish this Lagrangian from that of the MSSM with heavy fields not
replaced by their equations of motion)
\begin{eqnarray}
{\cal L}'_{MSSM}&=&({\cal D}_\mu \str)^{\alpha\dagger}({\cal D}^\mu
\str)^\alpha+({\cal D}_\mu \h)^{\dagger}({\cal D}^\mu
\h)-M_R^2|\str|^2-m^2|\h|^2\nn\\
&&-h_t^2s_\beta^2|\str|^2|\h|^2-{1\over 2}g_s^2\sum_a(\str^{\alpha
*}T^a_{\alpha\beta}\str^\beta)^2-{1\over 8}g^2\sum_a(\h^\dagger \sigma^a
\h)^2\nn\\
&&-{1\over 72}{g'}^2\left[4|\str|^2+3
c_{2\beta}|\h|^2\right]^2
+[h_ts_\beta Q_L^\alpha c\, t_R^{c\alpha}\cdot \h+{\mathrm h.c.}]\nn\\
&&
+h_t^2s_\beta^2|X_t|^2(\str^{\alpha *}\h)\cdot
P_{L}(\partial^2)
(\overline{\h}\str^\alpha)
+h_t^4s_\beta^2c_\beta^2|\str|^2
\h\cdot P_{A}(\partial^2)\overline{\h}|\str|^2
\nn\\
&&+h_t^2s_\beta^2g_s^2(T^a_{\alpha\beta}T^a_{\rho\eta}
-h_t^2\delta_{\rho\alpha}\delta_{\eta\beta})
|X_t|^2(\str^{\alpha *}\h)\cdot
P_{L}(\partial^2)
\str^{\rho *}\str^\eta
P_{L}(\partial^2)
(\overline{\h}\str^\beta)\nn\\
&&-h_t^4s_\beta^4|X_t|^2\left[\str^{\alpha *}\h\cdot
P_{L}(\partial^2)
\overline{\h}\right]\left[\h\cdot
P_{L}(\partial^2)
\overline{\h}\str^\alpha\right]\nn\\
&&-\left[h_t^4c_\beta^2s_\beta^2
X_tY_t^*(\str^{\alpha *}\h)\cdot
P_{L}(\partial^2)
\str^\alpha
P_{A}(\partial^2)
\overline{\h}|\str|^2+{\mathrm h.c.}\right]\nn\\
&&+h_t^4c_\beta^2s_\beta^2
|X_tY_t|^2(\str^{\alpha *}\h)\cdot
P_{L}(\partial^2)
\str^\alpha
P_{A}(\partial^2)
\str^{\beta *}
P_{L}(\partial^2)
\overline{\h}\str^\beta\nn\\
&&+{1\over 2}\left[h_t^2\str^{\alpha *}Q_{Li}^{\alpha T}
i{\overline{\sigma}^\mu}^T\partial_\mu P_{\tilde{H}}(\partial^2)(\str^\beta
Q_{Li}^{\beta *})+{\mathrm
h.c.}\right]\nn\\
&&-\left\{g_s^2T^a_{\alpha\beta}T^a_{\rho\eta}\str^{\alpha *}
t_R^{c\beta T}\left[M_G P_{G}(\partial^2) c\ t_R^{c \rho}\str^{\eta *}
+i\sigma_\mu\partial^\mu P_{G}(\partial^2)
t_R^{c\eta *}\str^{\rho}\right]+
{\mathrm h.c.}\right\}\nn\\
&&-h_t^2c_\beta^2P_{A}(\partial^2)(Q_{Li}^{\alpha\dagger}c t_R^{c\alpha
*})(Q_{Li}^{\beta T}c\ t_R^{c\beta})
+...
\label{Lmssmp}
\end{eqnarray}
where $g$ and $g'$ are the $SU(2)_L$ and $U(1)_Y$ coupling
constants respectively; $\overline{\h}$ is the $SU(2)$ conjugate of
the light Higgs field $\h$; $c_\beta=\cos\beta$, 
$s_\beta=\sin\beta$;  $\overline{\sigma}^\mu=(1,-\vec{\sigma})$; 
$T^a_{\alpha\beta}$ are the $SU(3)_C$ generators in
the fundamental representation; and 
\be
Y_t\equiv A_t-\mu^*\tan\beta\ .
\ee 
\begin{figure}[t]
\psfig{figure=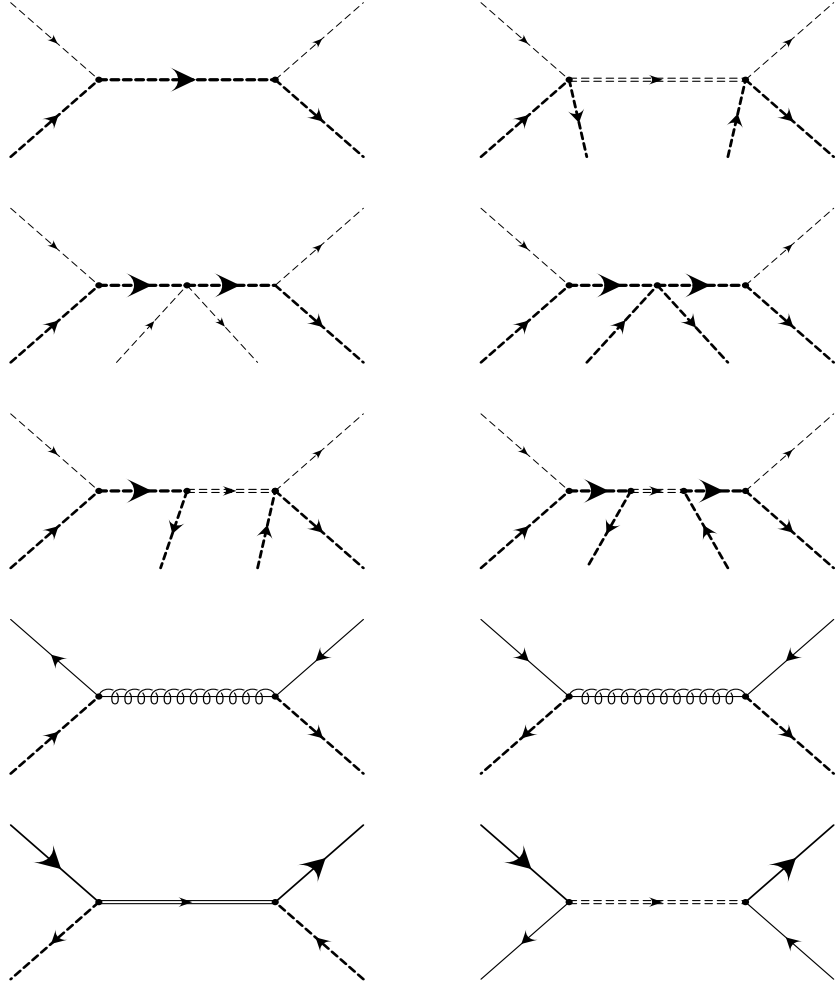,height=14cm,width=8cm,bbllx=-1.cm,%
bblly=-8.cm,bburx=12.cm,bbury=18.cm}
\caption
{\footnotesize
Tree-level diagrams, with heavy supersymmetric particles interchanged, that
give rise to different terms of the Lagrangian (\ref{Lmssmp}).}
\label{fig:Lag}
\end{figure}
\begin{figure}[t]
\psfig{figure=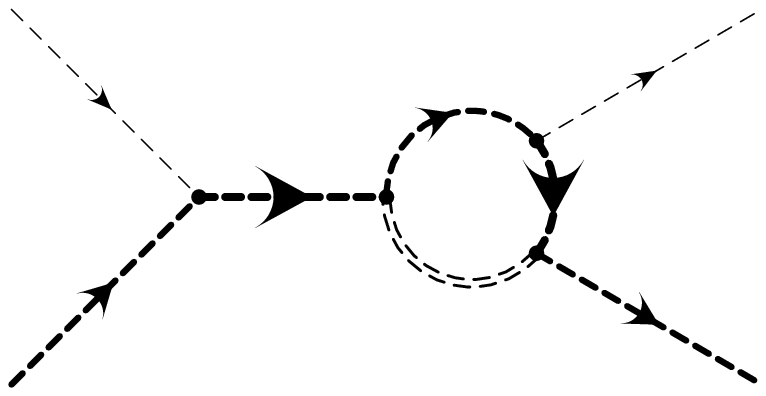,height=5cm,width=6cm,bbllx=-4.cm,%
bblly=-4.cm,bburx=4.cm,bbury=3.cm}
\caption
{\footnotesize
One-loop diagram that cancels when all the  components
of $SU(2)$ doublets run in the loop.}
\label{fig:can}
\end{figure}
We have written $\overline{M}_R^2$ and $\overline{m}^2$ in 
${\cal L}_{SM+\str}$ to distinguish them from ${M}_R^2$ and
${m}^2$ in ${\cal L}'_{MSSM}$. We have also
introduced the operators
\begin{eqnarray}
P_{x}(\partial^2)\equiv\left[{1\over
M_x^2+\partial^2}\right]_E\ ,
\label{op1}
\end{eqnarray}
for $x={L,A,\tilde{H},G}$ (corresponding to the soft mass $M_L$ of
$\tilde{t}_L$, the pseudoscalar mass $M_A$, the higgsino mass
$M_{\tilde{H}}^2=|\mu|^2$ and the gluino
mass $M_G^2$, respectively). In eq.~(\ref{op1}) the subindex
$E$ indicates a low-momentum expansion in powers of $\partial^2/M_x^2$. In
(\ref{Lmssmp}), these operators act only inside the square brackets they are
in. 
\begin{figure}[t]
\psfig{figure=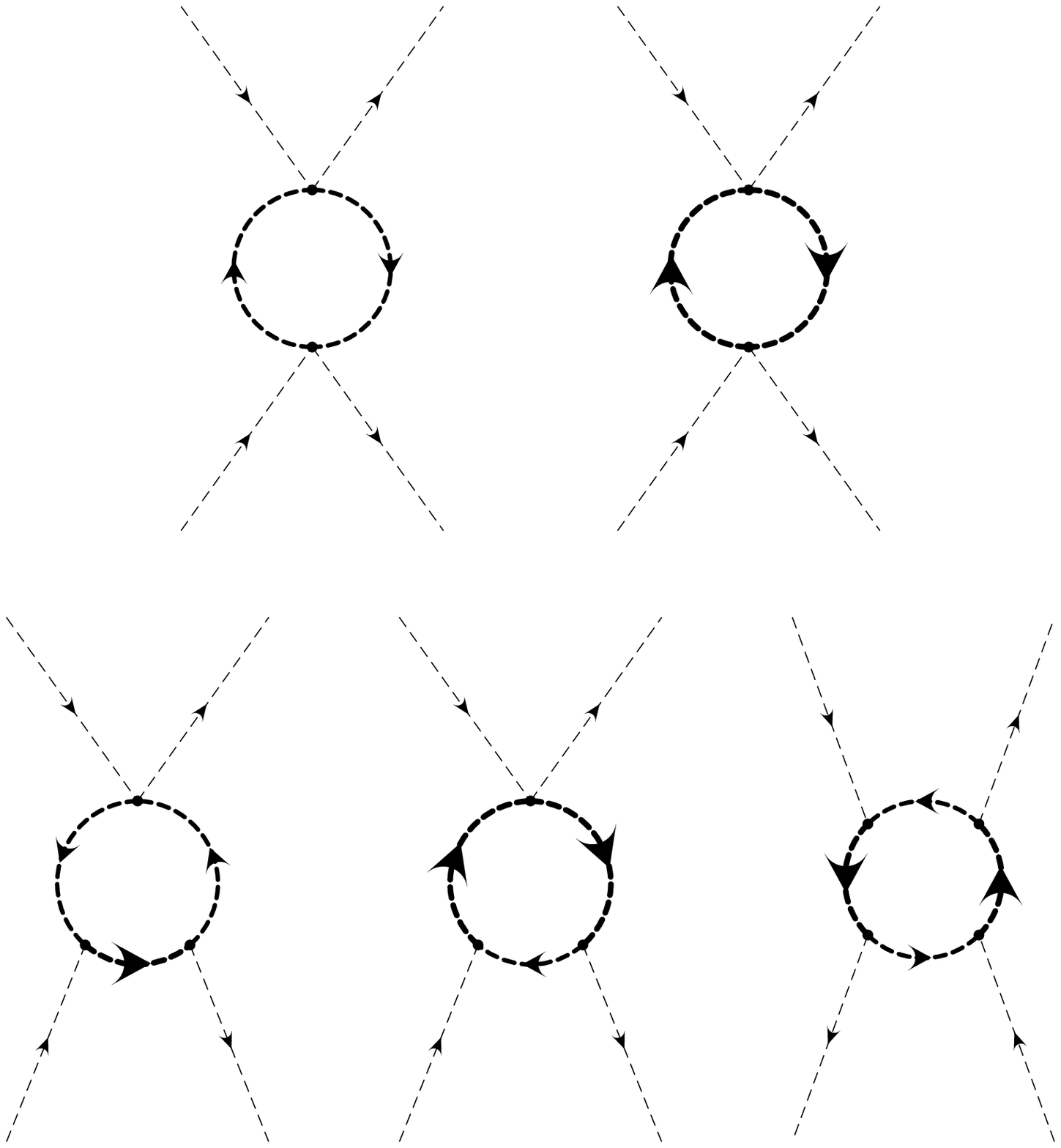,height=10cm,width=8cm,bbllx=-3.cm,%
bblly=-13.cm,bburx=10.cm,bbury=4.cm}
\caption
{\footnotesize
One-loop diagrams for the threshold correction to $\lambda_H$ in the
full MSSM.}
\label{fig:lH}
\end{figure}
\begin{figure}[t]
\psfig{figure=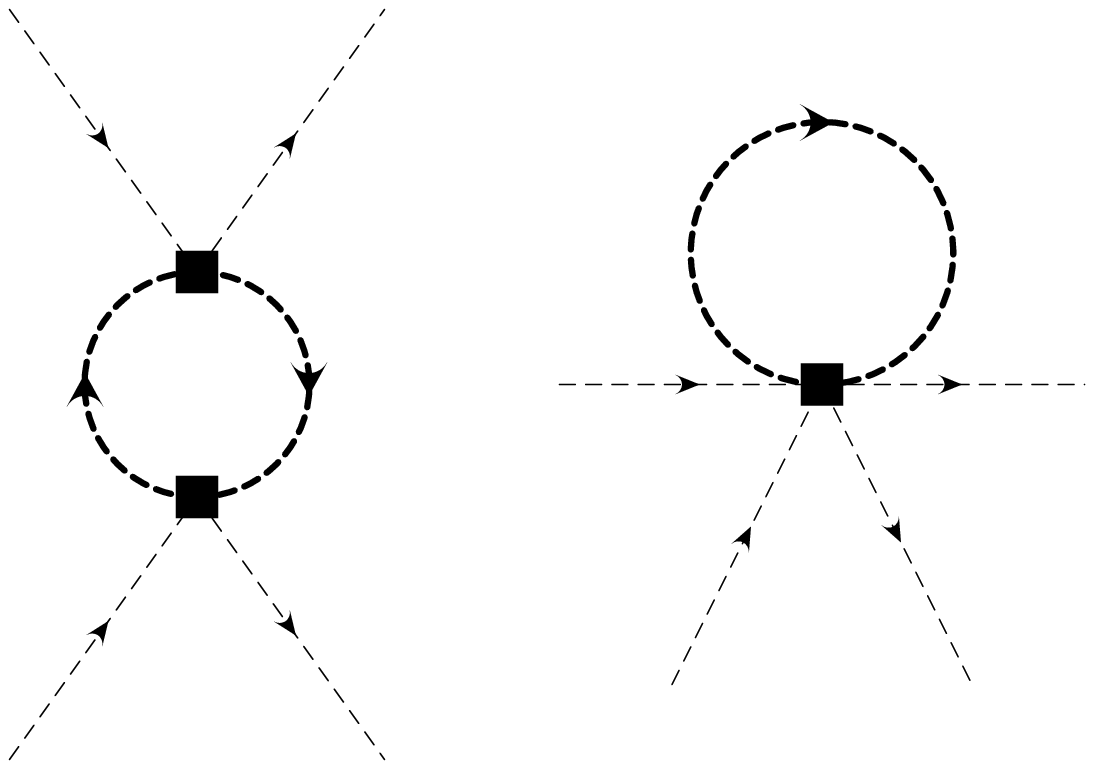,height=6cm,width=8cm,bbllx=-3.cm,%
bblly=-5.cm,bburx=10.cm,bbury=5.cm}
\caption
{\footnotesize
One-loop diagrams for the threshold correction to the quartic
Higgs coupling $\lambda_H$ in the
effective theory SM $+$ $\str$.}
\label{fig:lHeff}
\end{figure}

The origin of each non-renormalizable term in ${\cal L}'_{MSSM}$
[eq.~(\ref{Lmssmp})] is easy to
interpret as coming from the tree-level exchange of one or more heavy particles
[identified by the propagator operators $P_{x}(\partial^2)$]. This is shown
diagrammatically in figure~\ref{fig:Lag}, which shows the tree-level diagrams
that give
rise to the different terms of (\ref{Lmssmp}), upon collapsing heavy
particle lines to a point. The line code we use is the
following: a thin dashed line with a small arrow [which indicates the flow of
SU(2) quantum numbers] represents the light Higgs doublet; the same type of
line but with double dash corresponds to the heavy Higgs doublet; a double
continuous line with an $SU(2)$ arrow represents a Higgsino; a dashed bold
line with a large arrow [indicating $SU(3)$ colour flow] represents a light
stop; the same type of line but thicker and with a larger arrow [which
indicates the flow of $SU(2)$ and $SU(3)$ quantum numbers] is used for the
heavy stop; gluinos are represented by a continuous line with a wiggle; a
top-bottom quark doublet is represented by a solid line with a large arrow,
while the same type of line with a smaller arrow corresponds to a
right-handed top quark. For our calculations and diagrams we work
in the unbroken-symmetry phase, with the full $SU(2)$ doublet structure
unresolved. This simplifies our task: when dealing with separate diagrams for
the different components of a doublet there are cancellations, due to $SU(2)$
symmetry, which are immediately obvious in our approach. One such example is
the diagram of figure~\ref{fig:can}. There is a cancellation between the
contributions of 
different $SU(2)$ doublet components running in the loop. If one works with
complete $SU(2)$ multiplets, this cancellation is reflected in the
impossibility of drawing such a diagram properly: there is no choice
for the arrow of the heavy Higgs field in the loop that is consistent with
the flow of $SU(2)$ quantum numbers through other lines of the diagram.

Comparing ${\cal L}'_{MSSM}$ [eq.~(\ref{Lmssmp})] to ${\cal L}_{SM+\str}$
[eq.~(\ref{Lsmstr})], we get the tree-level matching
conditions:
\begin{eqnarray}
\lambda_H(M_L)&=&{1\over 4}(g^2+{g'}^2)\cos^22\beta
+\delta\lambda_H(M_L)\ ,\nn\\
\lambda_{HU}(M_L)&=&g_t^2-g_t^2{|X_t|^2\over 
M_L^2}+{1\over 3}{g'}^2\cos2\beta+\delta\lambda_{HU}(M_L)\ ,\nn\\
\lambda_U(M_L)&=&{1\over
3}g_s^2+{4\over 9}{g'}^2+\delta\lambda_U(M_L)\ ,\nn\\
g_t(M_L)&=&h_t\sin\beta+\delta g_t(M_L) \ , \nn\\
\overline{M}_R^2(M_L)&=&M_R^2+\delta \overline{M}_R^2(M_L)\ ,
\label{treematch}
\end{eqnarray}
where we have just indicated the presence of loop corrections and,
with an abuse of notation, used $g_t=h_t\sin\beta$ everywhere.
In addition, we list the following tree-level threshold values
for some non-renormalizable couplings in ${\cal
L}_{SM+\str}$ which will play a role in higher loop calculations:
\begin{eqnarray}
\lambda'_{HU}(M_L)&=&g_t^2{|X_t|^2\over M_L^4}+\delta\lambda'_{HU}(M_L)\
,\nn\\
\kappa^2(M_L)&=&g_t^4{|X_t|^2\over M_L^4}+\delta\kappa^2(M_L)\ ,\nn\\
\chi(M_L)&=&
\left({1\over 3}g_s^2-h_t^2\right){|X_t|^2\over M_L^4}+
{h_t^2c_\beta^2\over M_A^2}\left|1-{X_tY_t^*\over M_L^2}\right|^2
+\delta\chi(M_L)\ .
\label{moretreematch}
\end{eqnarray}

\begin{figure}[p]
\psfig{figure=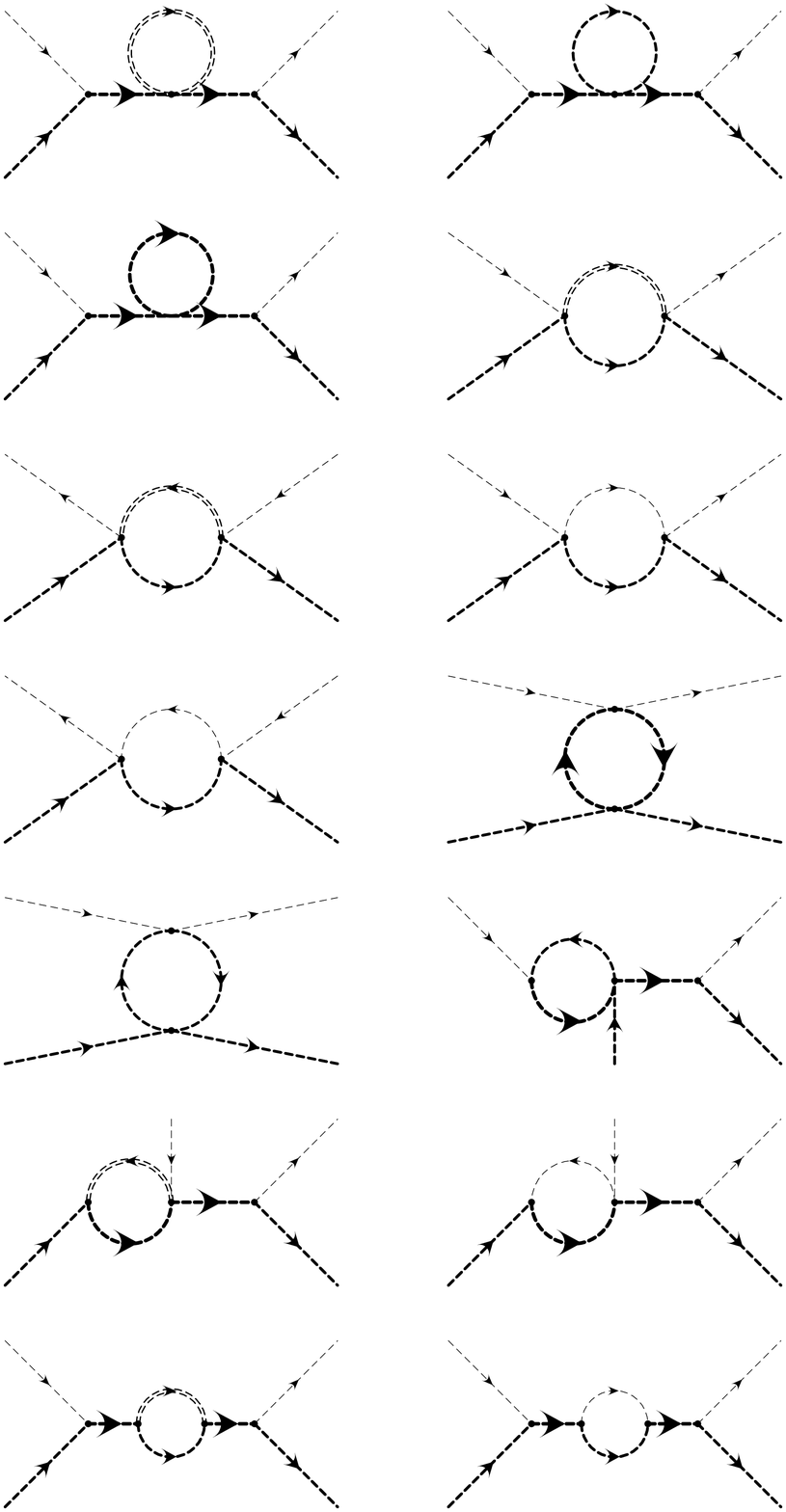,height=20cm,width=8cm,bbllx=-2.cm,%
bblly=-33.cm,bburx=12.cm,bbury=2.cm}
\caption
{\footnotesize
One-loop diagrams for the threshold correction to the Higgs-stop coupling 
$\lambda_{HU}$ in the full MSSM.}
\label{fig:lHU1}
\end{figure}
\begin{figure}[p]
\psfig{figure=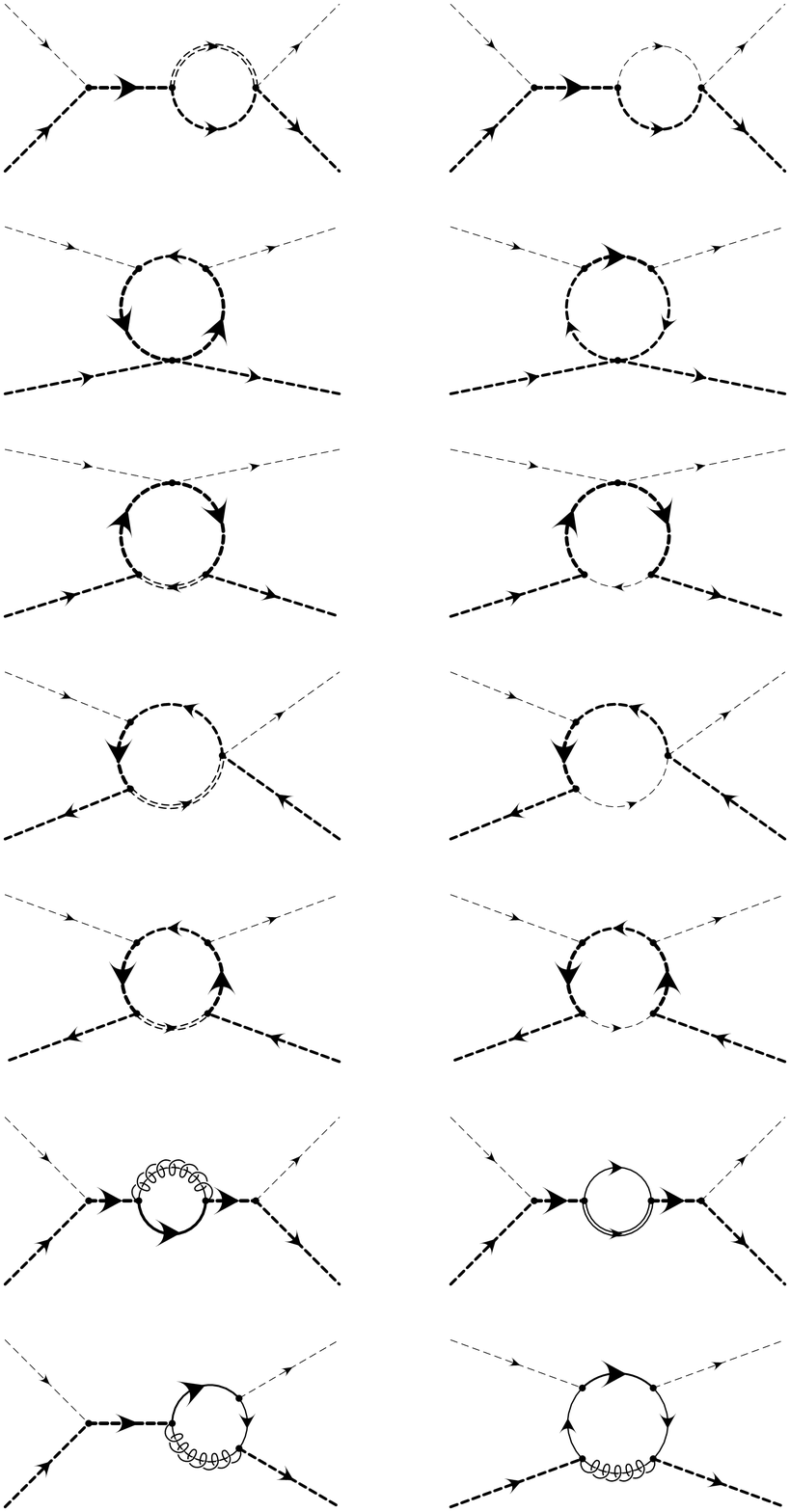,height=20cm,width=8cm,bbllx=-2.cm,%
bblly=-38.cm,bburx=12.cm,bbury=1.cm}
\caption
{\footnotesize
One-loop diagrams for the threshold correction to the Higgs-stop coupling
$\lambda_{HU}$ in the full MSSM (continued).}
\label{fig:lHU2}
\end{figure}
\begin{figure}[t]
\psfig{figure=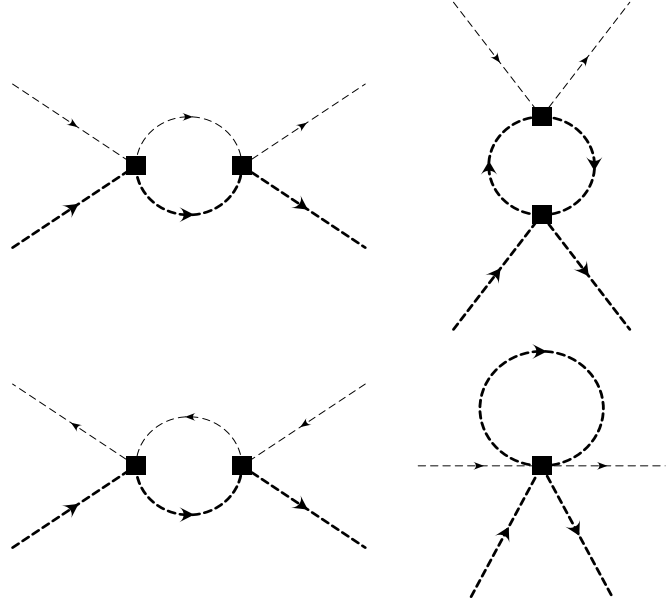,height=8cm,width=8cm,bbllx=-10.cm,%
bblly=-10.cm,bburx=3.cm,bbury=4.cm}
\caption
{\footnotesize
One-loop diagrams for the threshold correction to the Higgs-stop coupling  
$\lambda_{HU}$ in the effective theory SM $+$ $\str$.}
\label{fig:lHUeff}
\end{figure}

\begin{figure}[p]
\psfig{figure=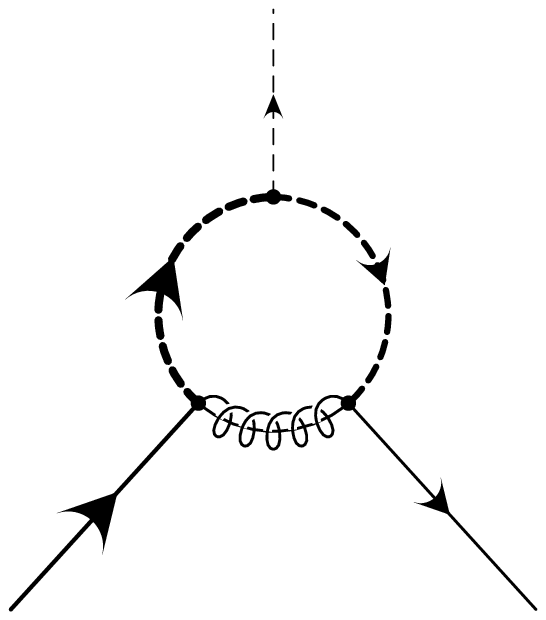,height=5cm,width=6cm,bbllx=-4.cm,%
bblly=-4.cm,bburx=4.cm,bbury=3.cm}
\caption
{\footnotesize
One-loop diagram for the threshold correction to the top Yukawa coupling 
in the full MSSM.}
\label{fig:gt}
\end{figure}

This tree-level matching would be enough if we were only after leading-log
corrections to $\mh$ (which are not sensitive to threshold corrections).  
As we have discussed already, if
we want to correctly obtain next-to-leading log contributions, this matching
must be done at one-loop level. That is, we need the one-loop threshold
corrections in the matching conditions (\ref{treematch}).  To compute them we
match the one-loop effective actions in both theories, MSSM and SM $+$
$\str$.
Again, to do this, we first substitute in the MSSM effective action the
equations of
motion of heavy fields in a local momentum expansion. In other words, we match
1LPI (one-light-particle-irreducible) graphs with light-particle external
legs in both theories. This procedure leads to the one loop threshold
corrections:
\begin{eqnarray}
\delta_0\lambda_H(M_L)&=&{N_c\over 8\pi^2}g_{t}^4{|X_t|^4\over
M_L^4}\left[1+3{M_R^2\over M_L^2}+{\cal O}\left({M_R^4\over
M_L^4}\right)\right]\ ,\\
\delta_0\lambda_{HU}(M_L)&=&{g_{t}^2\over 8\pi^2}\left\{
2g_s^2C_2(N_c)\left[1+2 M_G{X_t+h.c.\over M_L^2}\right]\right.\nn\\
&&+h_t^2\left[s_\beta^2+(1+2s_\beta^2+N_c){|X_t|^2\over    
M_L^2}+s_\beta^2{|X_t|^4\over M_L^4}
\right.\\
&&\left.
\left.+c_\beta^2{Y_tX_t^*+h.c.\over
M_L^2}+c_\beta^2{|Y_t|^2\over
4M_L^2}-3c_\beta^2{|X_tY_t|^2\over 4M_L^4}\right]\right\}+{\cal
O}\left({M_R^2\over M_L^2}\right)
\nn\ ,\\
\delta_0g_{t}(M_L)&=&-{g_{t}g_s^2\over8\pi^2} C_2(N_c)
{M_G\over M_L^2}X_t+{\cal O}\left({M_R^2\over M_L^2}\right)\ ,\\
\delta_0 M_R^2(M_L)&=&{g_s^2\over 4 \pi^2}C_2(N_c)M_S^2+{g_t^2\over
8\pi^2}(M_S^2-|X_t|^2)\ ,
\label{1lth}
\end{eqnarray}
Here, $N_c=3$ is the number
of colours and $C_2(N_c)=(N_c^2-1)/(2N_c)$ is the quadratic Casimir of the
fundamental representation of $SU(N_c)$.  For simplicity in our notation,
we remove the bar from $\overline{M}_R$ in these equations and from
now on, in the understanding that $M_R$ is, in what follows, the parameter of
the intermediate-energy theory.

These results are approximations in which we have neglected all
couplings other than the top Yukawa coupling and the strong gauge coupling. 
They are expansions in powers of $M_R/M_L$ to the indicated order of
approximation. We have truncated the expansions in such a way as to
reproduce the correct one-loop corrections to $\mh$ up to and including ${\cal
O}(M_R^2/M_L^2)$ terms while we neglect this type of corrections in the
two-loop contributions [for that reason we need to keep ${\cal
O}(M_R^2/M_L^2)$ corrections only to $\lambda_H$, which enters $\mh$ already
at tree level]. In addition, we have used the assumed identity of different
heavy masses ($M_A=M_{\tilde{H}}=M_G=M_L$) to simplify the expressions
(although we leave explicit linear terms in $M_G$ to allow for possible sign
effects related to that mass).

In figure~\ref{fig:lH} we give the 1LPI diagrams that contribute to
$\lambda_H$ at one loop in the full MSSM, while figure~\ref{fig:lHeff} shows
the corresponding diagrams in the SM $+$ $\str$ theory. Couplings in this last
figure are distinguished from those in figure~\ref{fig:lH} by a black square
to represent that they already include tree-level matching corrections. The
one-loop threshold correcction $\delta\lambda_H$ is given by the contribution
of the diagrams of figure~\ref{fig:lH} minus the contribution of the diagrams
of figure~\ref{fig:lHeff}. We therefore omit from these figures those diagrams
that would be exactly equal in both theories (such diagrams do not contribute
to the threshold correcction $\delta\lambda_H$) or diagrams that are simply
zero.

\begin{figure}[p]
\psfig{figure=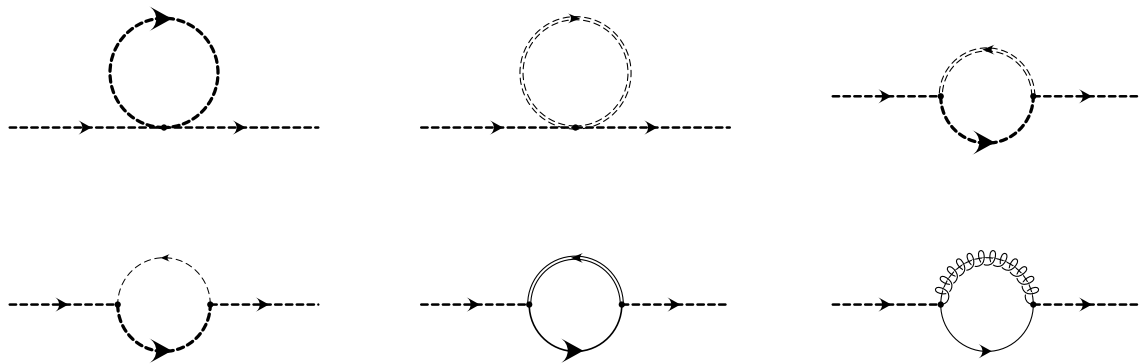,height=6cm,width=7cm,bbllx=2.cm,%
bblly=-7.cm,bburx=15.cm,bbury=4.cm}
\caption
{\footnotesize
One-loop diagrams for the threshold correction to the mass and
wave-function of $\str$ in the full MSSM.}
\label{fig:MR}
\end{figure}

In a similar way, figures~\ref{fig:lHU1} and \ref{fig:lHU2} give the 1LPI
diagrams that
contribute to $\lambda_{HU}$ at one loop in the full MSSM, while
figure~\ref{fig:lHUeff}
shows the
corresponding 1LPI diagrams in the SM $+$ $\str$ theory. 
At the order we work, the only
diagrams that contribute to the threshold corrections of
the top Yukawa coupling, $g_t$, and of the mass $M_R$ of the light stop
$\str$, are diagrams in the full MSSM. They are shown in figure~\ref{fig:gt} 
for
$g_t$, and in figure~\ref{fig:MR} for $M_R$.

\begin{figure}[p]
\psfig{figure=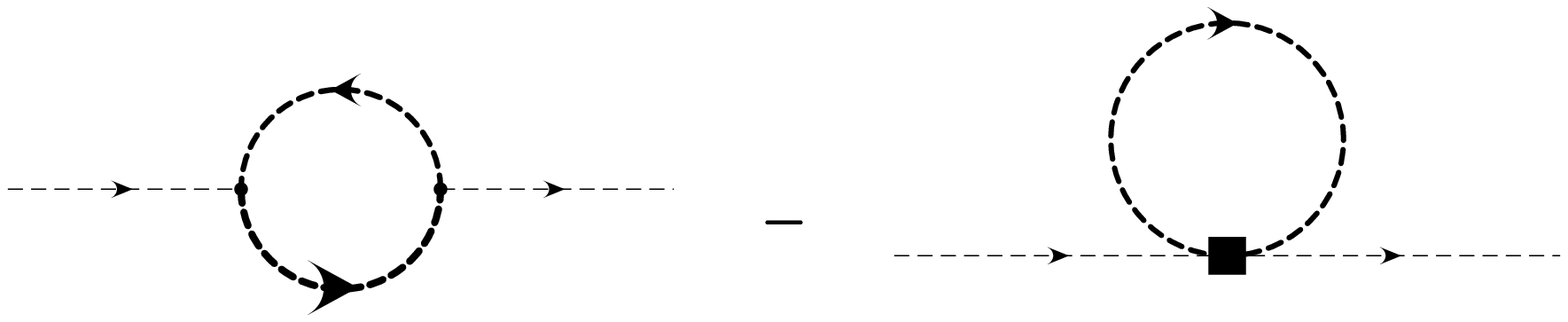,height=6cm,width=7cm,bbllx=-2.cm,%
bblly=-4.cm,bburx=11.cm,bbury=7.cm}
\caption
{\footnotesize
One-loop diagrams for the threshold correction to the wave-function
of the light Higgs $\h$ in the full MSSM and the effective theory
SM $+$ $\str$.}
\label{fig:ZH}
\end{figure}
\begin{figure}[t]
\psfig{figure=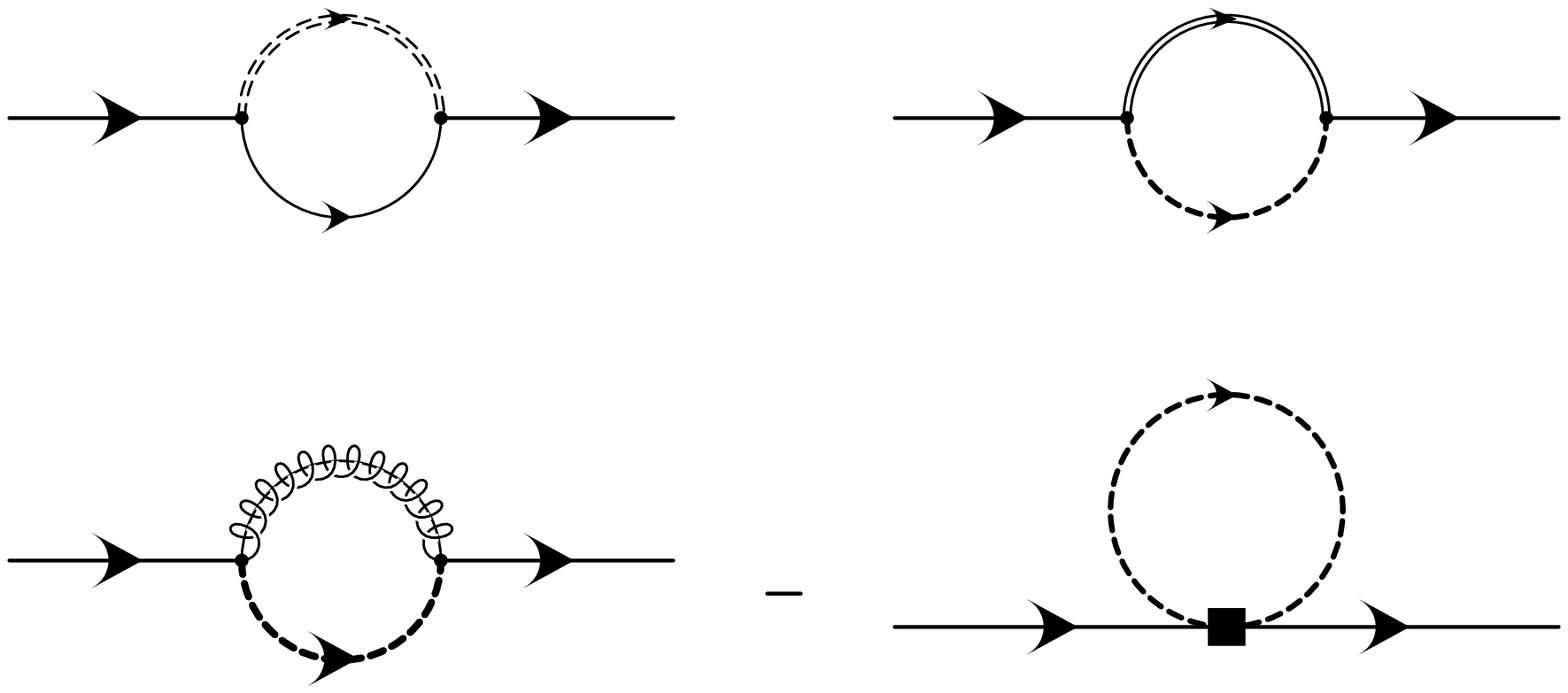,height=6cm,width=8cm,bbllx=-1.cm,%
bblly=-7.cm,bburx=12.cm,bbury=3.cm}
\caption
{\footnotesize
One-loop diagrams for the threshold correction to the wave-function 
of the left-handed top quark in the full MSSM and the effective theory 
SM $+$ $\str$.}
\label{fig:ZtL}
\end{figure}
\begin{figure}[t]
\psfig{figure=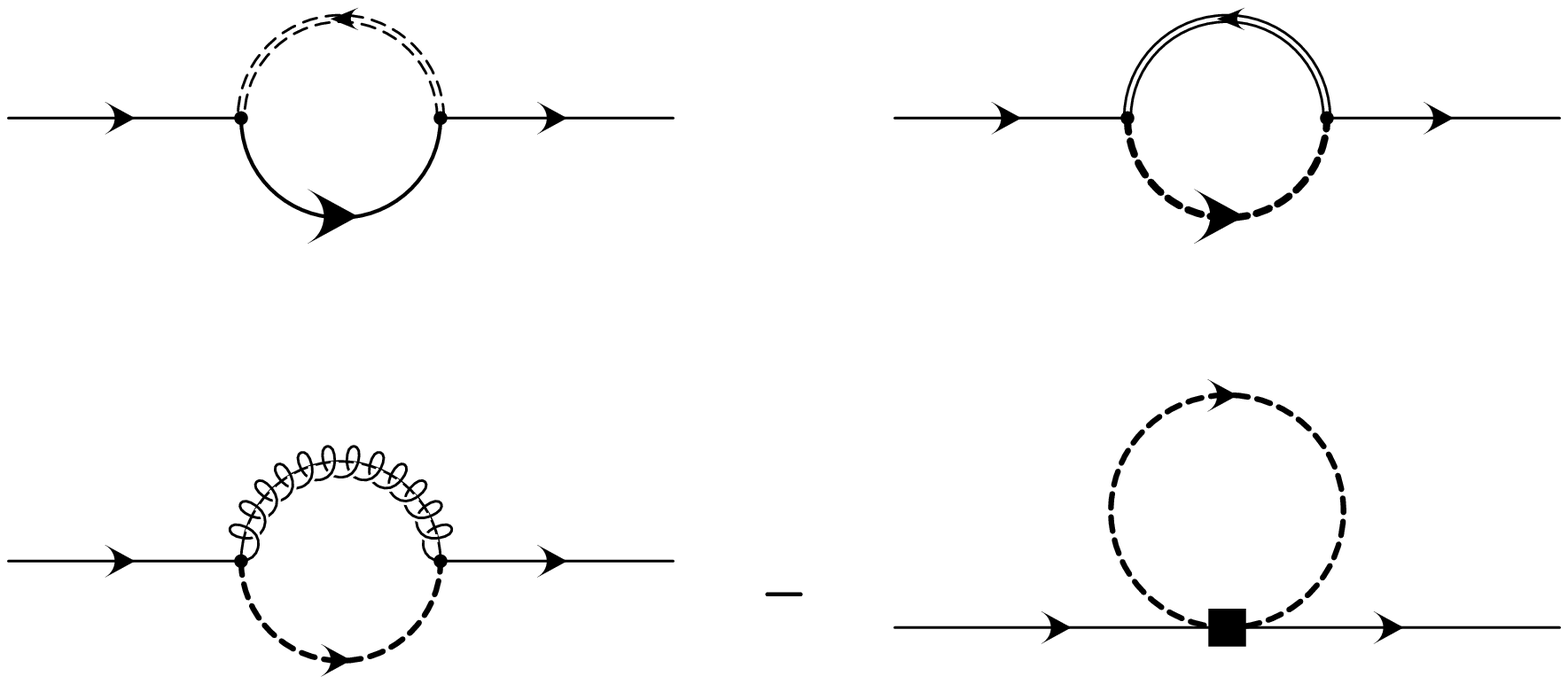,height=6cm,width=8cm,bbllx=-1.cm,%
bblly=-7.cm,bburx=12.cm,bbury=3.cm}
\caption
{\footnotesize
One-loop diagrams for the threshold correction to the  wave-function
of the right-handed top quark in the full MSSM and the effective theory 
SM $+$ $\str$.}
\label{fig:ZtR}
\end{figure}

The subindex $0$ in eq.~(\ref{1lth}) is meant to indicate that 
these are not the final one-loop threshold corrections:
there are also threshold corrections to kinetic terms and, after
redefining the fields to get canonical kinetic terms, the 
results in eq.~(\ref{1lth}) are also affected and one finally gets 
\begin{eqnarray}
\delta_1\lambda_H(M_L)&=&\delta_0\lambda_H(M_L)-2\lambda_H\delta Z_H\
,\nn\\
\delta_1\lambda_{HU}(M_L)&=&\delta_0\lambda_{HU}(M_L)-\lambda_{HU}(\delta 
Z_U+\delta Z_H)\ ,\nn\\
\delta_1 g_t(M_L)&=&\delta_0 g_t(M_L)-{1\over 2}g_t\left(\delta
Z_{t_L}+\delta Z_{t_R}+\delta Z_H\right)+\Delta_{\dr\rightarrow\ms}\
,\nn\\
\delta_1 M_R^2(M_L)&=&\delta_0 M_R^2(M_L^2)- M_R^2\delta Z_U\ ,
\label{onethres}
\end{eqnarray}
with wave-function threshold corrections encoded in
\begin{eqnarray}
\delta Z_{t_L}&=&{h_t^2\over 64\pi^2}(3+c_\beta^2)\ ,\nn\\
\delta Z_{t_R}&=&{3g_s^2\over 32\pi^2}C_2(N_c)+{h_t^2\over
32\pi^2}c_\beta^2\ ,\nn\\
\delta Z_H&=&{3 g_t^2\over 32 \pi^2}|X_t|^2\ ,\nn\\
\delta Z_U&=&{g_s^2\over 16\pi^2}C_2(N_c)+{h_t^2\over
16\pi^2}
\left(1+{|X_t|^2\over M_L^2}s_\beta^2+{1\over
3}{|Y_t|^2\over M_L^2}c_\beta^2\right)\ ,
\label{dZ}
\end{eqnarray}
for $Q_L^\alpha$, $t_R^\alpha$, $\h$ and $\str$ fields, respectively.
We have also added a term (see \cite{martin})
\be
\Delta_{\dr\rightarrow\ms}={g_{t}g_s^2\over16\pi^2}C_2(N_c) \ ,
\label{drms}\\
\ee
to correct from the change of scheme, from $\dr'$ (the modified $\dr$
scheme of \cite{DR}) in the MSSM to $\ms$, which is the scheme we use below
the supersymmetric threshold.

The diagrams that contribute to the threshold corrections to kinetic terms,
given in (\ref{dZ}), are shown in figure~\ref{fig:MR}, for $\str$; in
figure~\ref{fig:ZH} for $\h$; in figure~\ref{fig:ZtL} for $t_L$ and in
figure~\ref{fig:ZtR} for $t_R$. In the last three figures we give together the
diagrams in the full theory and (with a minus sign in front) those in the
effective theory (SM $+$ $\str$).

Several comments on the one-loop threshold corrections we have presented are
in order. To get the correct matching conditions it is important that both
heavy and light particles propagate in loops when computing the full MSSM
effective action (see {\it e.g.} the discussion in \cite{nyffeler}). We have
computed the above threshold corrections evaluating the functional determinant
expression for the effective action and also by direct diagrammatic
calculation of the matched graphs (for this task, the general reference
\cite{aoki} was helpful, as usual). The former method has the advantage of
being
systematic and of simplifying the determination of symmetry factors, the
second illuminates the physical origin of what is being computed. We find
agreement between the results from both methods.

As expected on general grounds \cite{Efft}, the threshold corrections are
infrared
finite, {\it i.e.} all sigularities in the limit $M_R\rightarrow 0$ cancel
in the matching. This happens for the threshold corrections and for all
their derivatives with respect to the light mass $M_R$. In particular, no
dependence on $\ln M_R$ is left in threshold corrections. As is well
known, for this cancellation of infrared divergences to occur, it is
crucial to keep enough derivatives in the low-energy effective couplings of
${\cal L}_{SM+\str}$ [eq.~(\ref{Lsmstr})]. This successful cancellation
provides a partial check of our results.

\begin{figure}[t]
\psfig{figure=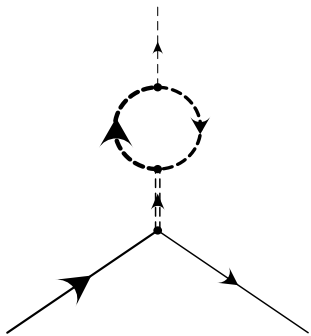,height=5cm,width=6cm,bbllx=-4.cm,%
bblly=-4.cm,bburx=4.cm,bbury=3.cm}
\caption
{\footnotesize
One-loop diagram for the threshold correction to the top Yukawa coupling 
in the full MSSM  with light-heavy Higgs doublet mixing.}
\label{fig:Hh}
\end{figure}

The treatment of the matching in the Higgs sector 
requires some
explanation. For large pseudoscalar mass, $M_{A^0}$, we can rotate the
original two Higgs
doublets of the MSSM, $\Hh_1$ and $\Hh_2$, into light and
heavy doublets ($\h$ and $\Hh$, respectively) which are given by
\be
\left(
\begin{array}{c}
\h \\[2mm]
\overline{\Hh}
\end{array}\right)
=\left(
\begin{array}{cc}
\cos\beta &-\sin\beta\\[2mm]
\sin\beta & \cos\beta
\end{array}
\right)
\left(
\begin{array}{c}
-\overline{\Hh}_1 \\[2mm]
\Hh_2
\end{array}
\right)
\ee
At one-loop order, however, a mixing between $\h$ and $\Hh$ is induced  
({\it e.g.} by stop loops). As we would like our field $\h$ to be the true
light Higgs doublet at one-loop, we treat the Higgs sector at one-loop in
the matching of effective theories. More precisely, we use one-loop
equations of motion for the heavy Higgs field. If this is done, all 1LPI
diagrams in the full theory that contain a external leg in which $\Hh$
switches to $\h$ through a loop correction, cancel exactly with the
one-loop contribution from the equation of motion for $\Hh$. This is what
one expects of Higgs fields properly diagonalized at one-loop level.
As a result, such diagrams do not contribute to one-loop threshold
corrections and, for that reason, we have not included them in previous
figures. An example of such diagrams is given in figure~\ref{fig:Hh} for the
top
Yukawa 
coupling.

\subsection{Running down to $M_R$}

Once we have fixed the couplings of the intermediate-energy theory at the
scale $Q=M_{SUSY}$ in terms of the parameters of the full theory, we run
them down in energy until we reach the next threshold at $M_R$. The
renormalization group equations (RGEs) in the intermediate effective
theory described by ${\cal L}_{SM+\str}$ [eq.~(\ref{Lsmstr})] can be
easily computed at one-loop ({\it e.g.} through the effective action)
including where necessary the effect of non-renormalizable operators. For
two-loop results we particularize to this theory the general formulas
presented in \cite{RGE2L}.

Following the notation introduced in eq.~(\ref{Lsmstr}) for the couplings of
the intermediate-energy theory, we have, for the
Higgs quartic coupling:
\begin{eqnarray}
\beta_{\lambda_H}^{(1)}&\simeq&{N_c\over 16\pi^2}\left[2\kappa^2
M_R^2+\lambda_{HU}^2-4 \lambda_{HU}{\lambda_{HU}'}M_R^2-2g_t^4\right]-
2\lambda_H\gamma_H^{(1)}\ ,\nn\\
\beta_{\lambda_H}^{(2)}&\simeq&{4N_c\over (16\pi^2)^2}\left[{5\over 2}
g_{t}^6+2g_s^2C_2(N_c)(\lambda_{HU}^2-g_{t}^4)-\lambda_{HU}^3
\right]\ ,
\label{RGEsSMtR}
\end{eqnarray}
with
\be
\gamma_H^{(1)}=-N_c{g_{t}^2\over 16\pi^2}\ ,
\label{gammaH}
\ee
describing the wave-function renormalization of the Higgs field.

In view of eq.~(\ref{plan}), we need to compute
$d\beta_{\lambda_H}^{(1)}/d\ln Q^2$ and, therefore, we also need the RGEs
for $\lambda_{HU}$ and $g_t$ (albeit only at one-loop). Those for $\kappa^2$
and $\lambda_{HU}'$ will not be needed because of the additional factor
$M_R^2$ which we neglect in two-loop corrections. For the Higgs-stop
coupling we find
\be
\beta_{\lambda_{HU}}^{(1)}\simeq {1\over 16\pi^2}\lambda_{HU}\left[(N_c
+1)\lambda_U+2\lambda_{HU}+3\lambda_H\right]
-\lambda_{HU}(\gamma_H^{(1)}+\gamma_U^{(1)})\ ,
\ee
with
\be
\gamma_U^{(1)}=(3-\alpha){g_s^2\over 16\pi^2}C_2(N_c)\ ,
\ee
describing the wave-function renormalization of the stop field (in
Landau gauge, in which we work, $\alpha=0$). For the top-Yukawa coupling we
get
\be
\beta_{g_{t}^2}^{(1)}=\frac{d g_{t}^2}{d\ln Q^2}\simeq {g_{t}^2\over
16\pi^2}\left({9\over 2}g_{t}^2-8 g_s^2\right)\ .
\label{bgt2}
\ee
Finally, we also give
\be
\beta_{M_R^2}^{(1)}=\frac{d M_R^2}{d\ln Q^2}\simeq  -{g_s^2\over
6\pi^2}M_R^2\ ,
\label{RGtop}
\ee
which will be useful later on to make contact with effective potential
results. We remind the reader that $M_R(Q)$ is the running mass of $\str$
in the intermediate effective theory, not in the full MSSM.

These RGEs are approximations in which the $SU(2)_L$ and $U(1)_Y$ gauge
couplings and all Yukawa couplings other than $g_t$ are neglected. We keep
$\lambda_H$ in (\ref{RGEsSMtR}) because it is formally of order
$g_t^4/(16\pi^2)$ and, therefore, the $\lambda_H\gamma_H^{(1)}$ term in
$\beta_{\lambda_H}^{(1)}$ is of the same order of $\beta_{\lambda_H}^{(2)}$
and should be kept\footnote{In fact, $\lambda_H\gamma_H^{(1)}$  is even
dominant with respect to $\beta_{\lambda_H}^{(2)}$ because of 
the logarithmic enhancement $\lambda_H\sim 
g_t^4/(16\pi^2)\ln(M_L^2/m_t^2)$. This term must be included even if we were
only interested in leading-log corrections \cite{CEQW}.}.

In $\beta_{\lambda_H}^{(1)}$ we also keep the contributions from
non-renormalizable operators ($\kappa^2$ and $\lambda_{HU}'$) which are
suppressed by a $M_R^2/M_L^2$ factor. This is necessary if we want that
precision in one loop corrections to $\mh$. At two-loops we do not keep such
subdominant terms and, consequently,  such contributions have not been kept in 
$\beta_{\lambda_H}^{(2)}$ or $\beta_{\lambda_{HU}}^{(1)}$.

\subsection{Intermediate threshold: matching SM
$+$ $\tilde{t}_R$ with SM}

With the RGEs presented in the previous section we can then run the parameters
down to $M_R$. At $M_R$ we need to match the low-energy effective theory, which
is the SM (with non-renormalizable operators formed out of SM fields):
\be
{\cal L}_{SM}=
({\cal D}_\mu \h)^{\dagger}({\cal D}^\mu
\h)-m_h^2|\h|^2-{1\over 2}\lambda |\h|^4
+[g_tQ_L^\alpha c\, t_R^{c\alpha}\cdot \h+{\mathrm h.c.}]
+... \ ,
\label{Lsm}
\ee
to the intermediate effective theory, SM $+$ $\str$, given by the
Lagrangian (\ref{Lsmstr}) but with the heavy field $\str$ removed. Due to
the fact that no low-energy SM fields (or combinations of them) have the
quantum numbers of $\str^\alpha$, there are no threshold corrections at
tree level to the couplings in (\ref{Lsm}). At one-loop we find no 
threshold correction for the top Yukawa coupling, $g_t$, and a non-zero
correction to the Higgs quartic coupling:
\be
\lambda(M_R)=\lambda_H(M_R)+\delta\lambda(M_R)\ ,
\ee
with
\begin{eqnarray}
\delta\lambda(M_R)&=&-{N_c\over
8\pi^2}\left[\kappa^2(M_R^2)-\lambda_{HU}(M_R^2)\lambda'_{HU}(M_R^2)
\right]
M_R^2+{\cal O}\left({M_R^4\over M_L^4}\right)\nn\\
&=&
-{N_c\over 8\pi^2}g_{t}^4{|X_t|^4\over
M_L^4}{M_R^2\over M_L^2}+{\cal O}\left({M_R^4\over M_L^4}\right) .
\label{dlMR}
\end{eqnarray}
The diagrams that contribute to this correction from the intermediate theory
are those already depicted in figure~\ref{fig:lHeff}. In (\ref{dlMR}), the
couplings that
appear in the first expression should in principle be evaluated at the scale
$M_R$. At the order we work, however, the choice of that scale is not
important and we can simply use their tree-level values as computed at $M_S$
and given in eqs.~(\ref{treematch}) and (\ref{moretreematch}).  In this way we
get the final expression in terms of SUSY parameters. The difference between
this expression and the proper one is a two-loop next-to-leading-log term of
order $M_R^2/M_L^2$ and we neglect such terms.

\subsection{Running down to $m_t$}

Next we use the SM RGEs to run $\lambda$ and $g_t$ from $M_R$ down 
to the electroweak scale, say to $Q=m_t$. These RGEs are the following.
For the Higgs quartic coupling, $\lambda$ [normalized to $\mh^2=
\lambda v^2$, with $v=246$ GeV] we have
\begin{eqnarray}
\beta_{\lambda}^{(1)}&\simeq &-{2N_cg_{t}^4\over 16\pi^2} -2\lambda
\gamma_H^{(1)}\ ,\nn\\
\beta_{\lambda}^{(2)}&\simeq &{N_cg_{t}^4\over (16\pi^2)^2}\left[10
g_{t}^2-8 C_2(N_c)g_s^2\right]\ ,\label{RGEsSM}
\end{eqnarray}
with $\gamma_{H}^{(1)}$ as given in eq.~(\ref{gammaH}). We again keep the
term $\lambda\gamma_H^{(1)}$ in order to get the correct two-loop result
for $\mh$, but neglect all couplings other than $g_t$ or $g_s$. For the top
Yukawa coupling we get the same one-loop running as in the intermediate
theory, eq.~(\ref{bgt2}). Notice that the RGEs for the intermediate
theory, eqs.~(\ref{RGEsSMtR}), reduce to eqs.~(\ref{RGEsSM}) in the limit
$\kappa^2,\lambda_{HU}\rightarrow 0$, as they should.

\subsection{Mass formula}

Once we have $\lambda(m_t)$ we simply use the SM relation
\be
M_{h^0}^2=\left(1+{g_{t}^2\over 8\pi^2}\right)
\lambda(m_t)\overline{v}^2(m_t)\ ,
\label{lmh}
\ee
to obtain the Higgs mass. Here $\overline{v}$ is the Higgs vev, running with
RGE
\be
{d\overline{v}^2\over d\ln Q^2}=\gamma_H\overline{v}^2\ ,
\ee
with $\gamma_H$ as given, at one loop, in (\ref{gammaH}). The one-loop
correction factor in (\ref{lmh}) takes care of Higgs wave-function
renormalization effects \cite{SZ}. The fact that $\lambda$ itself is to be
considered of one-loop order makes it unnecessary to refine
eq.~(\ref{lmh}) with two-loop effects. 

If we use the results of previous subsections, we can write a more explicit
formula for $\mh$ using for $\lambda(m_t)$ in (\ref{lmh}) the expression
\begin{eqnarray}
\lambda(m_t)&\simeq&
\lambda_H^0(M_L)+\left\{\delta_1\lambda_H(M_S)+{3t_{LR}\over 16\pi^2}\left[
2g_{t}^4(M_R)-{\lambda_{HU}^0}^2(M_L)\right]\right\}[1+2\gamma_H^{(1)}t_L]\nn\\
&&+\left[\delta_1\lambda(M_R)+{6t_R \over
16\pi^2}g_{t}^4(m_t)\right] [1+2\gamma_H^{(1)}t_R]
+\delta_2\lambda_H-\beta_{\lambda_H}^{(2)}t_{LR}\nn\\
&&-{t_{LR}\over 16\pi^2}(\kappa^2-12
\lambda_{HU}\lambda_{HU}')M_R^2
-{6\lambda_{HU}\over 16\pi^2}\delta_1\lambda_{HU}(M_L)t_{LR}
-\beta_{\lambda}^{(2)}t_R\nn\\
&&+\left[{3\over 16\pi^2}\left(2g_{t}^2\beta_{g_{t}^2}
+\lambda_{HU}\beta_{\lambda_{HU}}\right)+\gamma_H\beta_{\lambda_H}
\right]t_{LR}^2+\left({6g_{t}^2\over
16\pi^2}\beta_{g_{t}^2}+\gamma_H\beta_\lambda\right)t_R^2\ ,\nn \\ 
\label{formula}
\end{eqnarray}
with
\be
t_R\equiv\ln{M_R^2\over m_t^2}\ ,\;\;\; t_L\equiv\ln{M_L^2\over m_t^2}\
,\;\;\; t_{LR}\equiv\ln{M_L^2\over M_R^2}\ . 
\ee
Once again, let us remark that in this formula the couplings are running
$\ms$ parameters evaluated at the indicated scales (such scales are not
explicit when different choices amount to three-loop effects). In
particular, the mass parameters are defined by $M_R\equiv M_R(M_R)$,
$M_L\equiv M_L(M_L)$ and $m_t\equiv \overline{m}_t(m_t)$ (here we write
$\overline{m}_t$ for the SM running top mass). This is important to compare
\cite{EZ,CH3W2} with results in OS scheme. The connection to OS parameters
is dealt with in Appendix~C. 

To have a simple expression we have kept $g_t$, in eq.~(\ref{formula}),
evaluated at different scales in different terms. They are related to
$g_t(m_t)$ by
\begin{eqnarray}
g_{t}^2(M_R)&=&g_{t}^2(m_t)+\beta_{g_{t}^2}t_R\ ,\nn\\
g_{t}^2(M_L)&=&g_{t}^2(m_t)+\beta_{g_{t}^2}t_L
-\delta_1g_{t}^2(M_L)\ ,
\end{eqnarray}
with $\beta_{g_{t}^2}$ given in (\ref{bgt2}) and $\delta_1g_{t}^2(M_L)$ in
(\ref{onethres}).
We have also separated explicitly the two-loop threshold correction to
$\lambda_H$ as $\delta_2\lambda_H$. This correction can be most easily
extracted from the two-loop
effective potential and is presented in Appendix~B.

\section{One-loop `improved' formula?}

In the case of a single supersymmetric threshold, $M_S\gg m_t$, it was shown
in
\cite{EZ,EZ2} (following similar ideas already presented in \cite{H3}) that
two-loop logarithmic corrections to $\mh^2$ could be absorbed in a one-loop
expression with running parameters evaluated at judiciously chosen scales. The
compact formula obtained allowed a simple approximation for the Higgs mass.
Could a similar `improved' one-loop expression for $\mh^2$ be found for the
case with a hierarchical stop spectrum?

The derivation of the compact `improved' formula in \cite{EZ2} is best
understood in the RG approach. It starts with the expression
\be
\lambda(m_t)=\lambda(M_S)-\left[\beta_\lambda^{(1)}(m_t)+\beta_\lambda^{(2)}
\right]\ln{M_S^2\over m_t^2}-{1\over 2}{d\beta_\lambda^{(1)}\over d\ln
Q^2}
\ln^2{M_S^2\over m_t^2}+...
\label{start}
\ee
which is the version of (\ref{plan}) which applies to the degenerate case
$M_R=M_L=M_S$. As usual, given $\lambda(m_t)$, the Higgs mass is obtained by
(\ref{lmh}). The idea  is to use the
freedom in choosing the scales in 
$\beta_\lambda^{(1)}(m_t)\ln[M_S^2(M_S)/ m_t^2(m_t)]$, in (\ref{start}), and
$\overline{v}^2(m_t)$, in (\ref{lmh}), to absorb two-loop corrections.

The two-loop leading-log term in (\ref{start}),
\be
-{1\over 2}{d\beta_\lambda^{(1)}\over d\ln Q^2}
\ln^2{M_S^2\over m_t^2}\ ,
\label{twoll}
\ee
is easy to absorb by noting that
\be
-\beta_\lambda^{(1)}(m_t)\ln{M_S^2\over m_t^2}=
-\beta_\lambda^{(1)}(Q_t)\ln{M_S^2\over m_t^2}-
{d\beta_\lambda^{(1)}\over d\ln Q^2}\ln{M_S^2\over m_t^2}
\ln{m_t^2\over Q_t^2}+...\ ,
\label{qt}
\ee
so that, if we choose $Q_t^2=m_t M_S$, the last term in (\ref{qt})
cancels (\ref{twoll}).
This choice of scale was first advocated in \cite{H3} and later confirmed by
\cite{EZ,EZ2}.

If we also want to absorb the two-loop next-to-leading-log term in
(\ref{start})
\be
-\beta_\lambda^{(2)}\ln{M_S^2\over m_t^2}\ ,
\label{twolnl}
\ee
we have to choose appropriately the scales $Q_v$ and $Q_t'$ in
\be
\beta_\lambda^{(1)} \overline{v}^2(Q_v)\ln{M_S^2(M_S)\over m_t^2(Q_t')}\ ,
\ee
(which appears as a one-loop correction in $\mh^2$)
to cancel (\ref{twolnl}). That a successful choice exists at all results from
the happy accidental relation  between the RG functions of the Standard
Model:  
\be
\beta^{(2)}_\lambda\simeq {2\over 3}\beta^{(1)}_\lambda\left[
\gamma^{(1)}_H-{1\over g_t^2}\beta^{(1)}_{g_t^2}
\right]\ .
\label{happy}
\ee
This gives (see \cite{EZ2}\footnote{We have left out of the discussion the
complications associated to the fact that $\beta_\lambda^{(1)}$ contains a
piece $-\lambda\gamma_H$, which is formally of higher order. The effects of
including this term properly can be reabsorbed in the scales at which one
evaluates $\overline{v}^2$. The final result is as presented in
\cite{EZ2}.}) ${Q_t'}^3=M_S^2 m_t$ and $Q_v=m_t$. Further two-loop
next-to-leading-log corrections to $\mh$ associated with one-loop threshold
corrections [like those coming from
$\lambda_H(M_L)=\lambda_H^0(M_L)+\delta_1\lambda_H(M_L) +...$, if the
parameters were evaluated at $m_t$ instead of $M_L$] are trivial to absorb
(in the previous example, just by choosing the scale to be $M_L$ instead of
$m_t$).

The success of this program rests on two pillars. One is the existence of a
single threshold at $M_S$, which allows a simple absortion of two-loop
leading-log corrections. The second is relation (\ref{happy}). Note,
however, that there is nothing fundamental about this relation. In fact, it
is spoiled if one includes electroweak gauge couplings. Also, there is no
reason to expect that leading or next-to-leading corrections beyond two
loops will be given correctly by the advocated choice of scales. In
conclusion, the compact formula was simply a useful approximation for the
two-loop result, with no pretension to being fundamental.

In view of the above discussion, the derivation of a compact one-loop
`improved' approximation to $\mh^2$ in the case of a hierarchical stop
spectrum looks problematic. To begin with,
eq.~(\ref{happy}) holds for the running of the Higgs quartic coupling between
$M_R$ and $m_t$ but not between $M_L$ and $M_R$; that is, $\beta_{\lambda_H}$
does not satisfy (\ref{happy}). Moreover, there are further complications 
associated with threshold corrections at $M_R$ which were absent in the
case of degenerate stop masses. All this implies that one cannot absorb
two-loop next-to-leading-log corrections in any simple way.

For two-loop leading-log corrections there are no threshold complications but
the difficulties with two regimes of running (above or below $M_R$) remain.
It is certainly possible to absorb these corrections using a trick similar 
to that in (\ref{qt}) in two separate one-loop
terms
\be
-\beta_\lambda^{(1)}(Q_1)\ln{M_R^2\over
m_t^2}-\beta^{(1)}_{\lambda_H}(Q_2)
\ln{M_L^2\over M_R^2}\ ,
\label{hier1li}
\ee
with $Q_1^2=M_L m_t$ and $Q_2^2=M_R m_t$, but $\beta_{\lambda_H}$ is a
function
of couplings like $\lambda_{HU}$ which are not present in the low-energy 
theory and one would like to do better than that.
An attempt in that direction was made in \cite{H3}, which for this
type of spectrum advocates
the use of 
\be
-\beta_\lambda^{(1)}(Q_3)\ln{M_L M_R\over m_t^2}\ ,
\label{apH3}
\ee
with the scale $Q_3$ defined by\footnote{With some refinements like
$M_L\rightarrow M_{\sto}$ and $M_R\rightarrow M_{\stw}$ which, in our
approximations, do not affect the leading-log corrections we are discussing
here.}
\be
\ln{Q_3^2 \over m_t^2}\ln{M_L M_R\over m_t^2} =
\ln^2{M_L\over m_t}+\ln^2{M_R\over m_t}\ ,
\label{q3}
\ee
as the one-loop redefinition of scales which absorbs two-loop leading-log
corrections. In view of eq.~(\ref{hier1li}), for this approximation to be
successful, $\beta_{\lambda_H}^{(1)}$ has to be related to
$\beta_{\lambda}^{(1)}$ somehow. If we substitute $\lambda_{HU}$ in
$\beta^{(1)}_{\lambda_H}$ by its
threshold value (\ref{treematch}), neglecting $X_t$-dependent terms (these
could be absorbed eventually in $X_t$-dependent one-loop corrections) one
has
\be
\beta_{\lambda_H}^{(1)}\simeq {1\over 2}\beta_{\lambda}^{(1)}\ ,
\label{betap}
\ee 
which only holds at $Q=M_L$. Were this to imply 
\be
{d\beta_{\lambda_H}^{(1)}\over d\ln Q^2}\simeq
{1\over2}{d\beta_{\lambda}^{(1)}\over d\ln Q^2}\ ,
\label{dbetap}
\ee
then, a one-loop approximation of the form (\ref{apH3}) could be devised, and
in fact we would find a scale $Q_3$ given precisely by (\ref{q3}). However,
(\ref{betap}) does not imply (\ref{dbetap}), which in fact does not hold, as
can be checked from the RGEs given in the previous subsections.

The best we can do to absorb two-loop leading-log corrections in a
one-loop formula is to use the freedom in choosing the scales of both
$\overline{m}_t$ [or, equivalently $\beta^{(1)}_{\lambda}$] and
$\overline{v}^2$, although this only works for
the $X_t$-independent corrections. If we focus only in the $X_t=0$ case,
the one-loop `improved' formula we find is 
\be
\Delta\mh^2= {3 \overline{m}^4_t(Q_t)\over 4\pi^2\overline{v}^2(Q_v)}
\ln{M_R^2M_L^2\over m_t^4}\ ,
\label{oneli}
\ee
with the scales $Q_t$ and $Q_v$ given by
\be
\ln{Q_t^2 \over m_t^2}\ln{M_L M_R\over m_t^2} =
{1\over 3}\left[\ln^2{M_R\over m_t}+4\ln{M_R\over m_t}\ln{M_L\over
m_t}+\ln^2{M_L\over m_t}\right]\ , 
\ee
which differs from the $Q_3$ defined in (\ref{q3}), and
\be
\ln{Q_v^2 \over m_t^2}\ln{M_L M_R\over m_t^2} =
\ln^2{M_L\over M_R}\ .
\ee
Formula (\ref{oneli}) successfully reproduces the two-loop leading
logarithmic corrections to $\mh^2$ for $X_t=0$.

However, for the general case, with non-zero $X_t$, we conclude that there
is no simple way of absorbing two-loop leading and next-to-leading
logarithmic corrections to $\mh^2$ in the case of a hierarchical stop
spectrum. This does not preclude the possible existence of scale choices
which minimize two-loop corrections, but we do not try to find them in this
paper.

\section{Comparison with previous literature}

Several papers have studied  two-loop radiative corrections for
non-degenerate stop masses before. In the previous discussion, we have already
mentioned ref.~\cite{H3}, which performed a thorough study of radiative
corrections to $\mh$ using RG techniques to derive analytical
and one-loop `improved' approximations that take into account two-loop
leading logarithmic corrections to the Higgs boson mass. As we have shown,
our results do not agree with those of \cite{H3} in the case in which we
focus, with widely different stop masses. A possible reason for this
discrepancy has been advanced above. 

Similar studies were performed also in ref.~\cite{CQW}, that uses a
combination of effective potential and RG techniques to obtain the
two-loop leading-log corrections to $\mh$ working in the more general case
of arbitrary pseudoscalar mass $M_{A^0}$ (this complicates the analysis
because, for low $M_{A^0}$, the low energy effective theory is a two Higgs
doublet model). However, if we compare with the results of that paper for
the case $M_{A^0}=M_S$ we again find they disagree with our results, and
the reason is similar to the one already mentioned: although at one-loop
there is a simple relation between some RG-functions of the effective
and full theories, its derivatives with the scale (which are necessary to
get correctly the two-loop-leading-log terms) are more complicated than
the one-loop relations suggest.

Before proceeding with the comparison to other previous analyses, we can 
already draw some implications of our results. It is simple to show,
focusing in two-loop-leading-log corrections and zero stop mixing for
simplicity, that our $\mh^2$ is higher than the previous estimates 
just commented \cite{CQW,H3} by the amount
\be
\Delta\mh^2={m_t^4\over 8\pi^4v^2}\left({3g_t^2\over 16}+2g_s^2\right)
\ln^2{M_L^2\over M_R^2}\ .
\label{disc}
\ee 
This formula assumes that the one-loop result is expressed in terms of
$\overline{m}_t(m_t)$ and $\overline{v}(m_t)$. As expected, the
discrepancy (\ref{disc}) is larger the higher $M_L^2/M_R^2$ is,
disappears if $M_L\simeq M_R$ and is always positive. Numerically, it
increases $\mh$ up to $\sim 5$ GeV in the most extreme cases (say $M_L\simeq 
2$ TeV and $M_R\simeq m_t$). The first obvious implication of this 
upward shift of $\mh$ is for experimental analyses of SUSY Higgs searches
that made use of those previous mass calculations, whenever they were
applied to scenarios with widely split stop masses. The same implications
will also follow for theoretical studies in similar circumstances.

Some two-loop radiative corrections to $\mh$ (those which depend on the QCD
gauge coupling) have been computed also diagrammatically \cite{hollik}, so
we could make a partial check of our results. However, a complete expression
for the diagrammatic result, applicable when the diagonal stop masses ($M_L$
and $M_R$) are different, is too lengthy and has not been published. It
would be interesting to make this comparison\footnote{The agreement between
these two-loop diagrammatic results and the corresponding RG and/or
effective potential results was first shown in \cite{EZ}. Unfortunately, this
comparison had to be limited to the case of heavy and degenerate diagonal
stop masses, $M_L=M_R=M_{SUSY}\gg m_t$. (See also \cite{CH3W2})}.

Finally there are two-loop calculations of $\mh$ based on the use
of the MSSM effective potential. The first of them was the work
by Hempfling and Hoang in \cite{H2}, which computed this potential for 
$\sin\beta=1$ and zero stop-mixing, extracting from it the
radiative corrections to the Higgs mass. This work was extended later on
by Zhang in \cite{zhang}, which added the QCD two-loop corrections to
the effective potential for generic values of $\tan\beta$ and non-zero stop
mixing. Finally, ref.~\cite{EZ2} included also the top-Yukawa two-loop
contributions to the MSSM effective potential. This last calculation
agrees with previous effective potentials in the different limits in which
those apply, so that we will only discuss here the comparison of our RG
two-loop corrections with those that can be obtained from the effective
potential as presented in \cite{EZ2}.

The procedure used to get the Higgs mass from the MSSM two-loop effective
potential is similar to the one used and explained in \cite{EZ,EZ2}. We
first expand this potential, $V$ (which, for large $M_{A^0}$, is a function of
the light Higgs field
through its dependence on $m_t$ and the stop masses)  in powers of
$m_t^2/M_L^2$, $m_t^2/M_R^2$ and $M_R^2/M_L^2$. Next we compute
\be
\mh^2(0)={4m_t^4\over v^2}\left({d\over d m_t^2}\right)^2V\ ,
\label{mh0}
\ee
which gives an approximation to $\mh^2$ as the second derivative of the
potential (with respect to the Higgs field) at its minimum. Then,
to get the physical Higgs pole mass, we correct for non-zero external
momentum effects by adding to (\ref{mh0}) the quantity
\be
\Pi_{hh}(0)-\Pi_{hh}(\mh^2)\ ,
\ee
where $\Pi_{hh}(p^2)$ is the Higgs boson self-energy for external momentum
$p$. What results from this procedure is an expression for $\mh^2$ in terms
of scale dependent $\dr'$-running parameters. To make contact with our
results in this paper we have to convert the parameters that survive below
the SUSY threshold to the $\ms$ scheme, taking into account one-loop
threshold corrections. For example, $m_t(Q)$, $M_R(Q)$ and $v(Q)$ are
expressed in terms of its low-energy counterparts $\overline{m}_t(Q)$,
$\overline{M}_R(Q)$ and $\overline{v}(Q)$ (these relations are given in
Appendix~B). The final step is to express all running parameters in terms of
their values at the scales at which we evaluate them in our RG approach: 
$\overline{v}(m_t)$, $m_t=\overline{m}_t(m_t)$, $M_R=\overline{M}_R(M_R)$,
$M_L=M_L(M_L)$ and $X_t(M_L)$. After doing this, we find an expression
for $\mh^2$ which is manifestly independent of the renormalization scale $Q$
and which agrees exactly with our effective theory result (we give its
expression explicitly in Appendix~A). As a bonus, the effective potential
calculation gives also the two-loop non-logarithmic correction, presented in
Appendix~B.

This agreement is the best check of our results. It speaks greatly of the
power of effective potential techniques: the effective potential
$V$ has built-in all the structure of RG-functions and threshold
corrections that we had to compute afresh in the RG approach. Nevertheless
our effort was not wasted because this structure which remains buried in
$V$, needs to be made explicit to implement the resummation of logarithmic
corrections to all loops of the RG programme. When the hierarchy in the
stop masses is only moderate there is no need to resum large logarithms
and one can revert to the use of the plain effective potential, which
still gives correctly two-loop radiative contributions to $\mh$. In other
scenarios that may introduce large logarithms ({\it e.g} when stop mixing
effects are large and cause a significant splitting of the stop masses or
for more complicated patterns of SUSY particle masses) one should work out
the relevant effective theories and the corresponding RG-functions and
threshold corrections. Still, to two-loop order the results of such
calculations must agree with those coming from the effective potential in
the same regime of parameters.

Needless to say, it would be extremely useful to have the tools necessary
to dig up all this structure from the effective potential directly.
Studies on multi-scale effective potentials \cite{multiV}, and in
particular the recent proposal in \cite{multiV1}, are a first
promising step in that direction and it would be interesting to 
continue the development of such techniques.

\section{Conclusions and outlook}

The mass $\mh$ of the MSSM light Higgs boson receives sizeable radiative
corrections, the most important of which depend on the details of the stop
spectrum (masses and mixing). In this paper we have revisited the
calculation of the radiative corrections to $\mh$ in the case of a
hierarchical stop spectrum, $M_{\sto}\gg M_{\stw} \gg M_t$ with moderate
stop mixing. We have used an effective theory approach to identify in these
corrections non-logarithmic contributions (which can be interpreted as
threshold corrections at different energy scales) and logarithmic
contributions (which arise from renormalization group running of parameters
between different energy thresholds). We have performed this calculation
neglecting in radiative corrections all couplings other than the top Yukawa
coupling and the QCD strong gauge coupling. Within this approximation we
collected all radiative corrections to $\mh$ up to, and including, two-loop
terms. Our results correct previous calculations
of two-loop leading-log corrections to the Higgs mass that appeared in the
literature and are widely used \cite{CQW,H3}, while we find complete
agreement with the results of previous analyses based on effective
potential techniques \cite{EZ,EZ2}.
Numerically, we find that two-loop leading-log corrections increase 
$\mh$ by up to 5 GeV (in some cases with a large hierarchy of stop
masses) relative to the results computed in \cite{CQW,H3}. This has
obvious importance for the theoretical input used in experimental
analyses.

Beyond this comparison to previous findings, the results obtained can be
used as the starting point of a numerical evaluation of the Higgs mass which
can resum leading and next-to-leading logarithmic terms to all loops using
renormalization group techniques. This resummation is mandatory to get an
accurate calculation of $\mh$ if the hierarchy of masses in the stop sector
is large. We defer such numerical analyses to some future paper.

The present analysis can be extended in several ways. It is simple to add
radiative corrections to $\mh$ that depend on the bottom Yukawa coupling,
which may be important for large values of $\tan\beta$. Also, formulas
similar to the ones derived in this paper could be found for the case of a
light stop, with $M_{\stw} \sim m_t$. Other extensions of our
results include the study of the radiative corrections to $\mh$ when the
hierachy between stop masses is due to a very large value of the stop mixing
parameter, $X_t$ or the connection of effective theory methods with the
methods of multi-scale potentials developed in \cite{multiV,multiV1}.

\newpage

\section*{Appendix A: Explicit Higgs mass formula}
\setcounter{equation}{0}
\renewcommand{\theequation}{A.\arabic{equation}}

In this Appendix we write down the Higgs boson mass computed in the main
text, explicitly separating the radiative corrections as
\be
\mh^2=M_Z^2\cos^22\beta+\Delta_{1LL}\mh^2+\Delta_{1NLL}\mh^2
+\Delta_{2LL}\mh^2+\Delta_{2NLL}\mh^2+\Delta_{2NNLL}\mh^2\ ,
\ee
where $\Delta_{1LL}\mh^2$ contains the one-loop leading-log corrections,
$\Delta_{1NLL}\mh^2$ the one-loop next-to-leading-log corrections,
$\Delta_{2LL}\mh^2$ the two-loop leading-log contributions,
$\Delta_{2NLL}\mh^2$ the two-loop next-to-leading-log contributions, and,
finally, $\Delta_{2NNLL}\mh^2$, the next-to-next-to-leading-log (that is,
the non-logarithmic) corrections. For two-loop terms, this division depends
on the scales at which the parameters in one-loop contributions are
evaluated. We take $m_t=\overline{m}_t(m_t)$, $v=\overline{v}_t(m_t)$,
$M_R=M_R(M_R)$, $M_L=M_L(M_L)$, $X_t=X_t(M_L)$ and $Y_t=Y_t(M_L)$. Here 
$\overline{m}_t(Q)$, $\overline{v}(Q)$ and $M_R(Q)$ are $\ms$-running
parameters in the effective theories below $M_{SUSY}$, and $M_L(Q)$,
$X_t(Q)$, $Y_t(Q)$ are $\dr'$-running parameters in the full MSSM.
For simplicity we take the stop mixing parameters to be real in the
following expressions.

For the one-loop pieces we obtain the well known results
\be
\Delta_{1LL}\mh^2={3 m_t^4\over 4\pi^2v^2}\left\{
\ln{M_R^2M_L^2\over m_t^4}+\left[2\hat{X}_t^2\left(1+{M_R^2\over M_L^2}
\right)-\hat{X}_t^4\left(1+4{M_R^2\over M_L^2}
\right)\right]\ln{M_L^2\over M_R^2}
\right\}\ ,
\ee
and
\be
\Delta_{1NLL}\mh^2={3 m_t^4\over 2\pi^2v^2}\left(1+2{M_R^2\over
M_L^2}\right)\hat{X}_t^4\ ,
\ee
where we have kept up to ${\cal O}(M_R^2/M_L^2)$ terms and used
$\hat{X}_t=X_t/M_L$, $\hat{Y}_t=Y_t/M_L$.

For two-loop corrections we find
\begin{eqnarray}
\Delta_{2LL}\mh^2&=&-{\alpha_s m_t^4\over 2\pi^3v^2}\left\{
\ln^2{M_R^2\over m_t^2}+4\ln{M_R^2\over m_t^2}\ln{M_L^2\over m_t^2}
+\ln^2{M_L^2\over m_t^2}\right.\nn\\
&&\left.+(2\hat{X}_t^2-\hat{X}_t^4)\left[5\ln{M_L^2\over m_t^2}+
\ln{M_R^2\over m_t^2}
\right]\ln{M_L^2\over M_R^2}\right\}\nn\\
&&+{3\alpha_{t,SM} m_t^4\over 16\pi^3v^2}\left\{
2\ln^2{M_R^2\over m_t^2}+2\ln^2{M_L^2\over m_t^2}
-\ln{M_R^2\over m_t^2}\ln{M_L^2\over m_t^2}\right.\nn\\
&&\left.+3(2\hat{X}_t^2-\hat{X}_t^4)\ln{M_R^2\over m_t^2}\ln{M_L^2\over
M_R^2}+(3\hat{X}_t^4-2\hat{X}_t^6)\ln^2{M_L^2\over M_R^2}\right\}\ ,
\end{eqnarray}
with $\alpha_s=g_s^2/(4\pi)$, $\alpha_{t,SM}=g_t^2/(4\pi)$; and
\begin{eqnarray}
\Delta_{2NLL}\mh^2&=&{\alpha_s m_t^4\over 4\pi^3v^2}\left\{
13 \ln{M_R^2\over m_t^2}-5\ln{M_L^2\over m_t^2}-24\hat{X}_t
\ln{M_L^2\over M_R^2}+18\hat{X}_t^2\ln{M_L^2\over M_R^2}\right.\nn\\
&&\left.+32\hat{X}_t^3\ln{M_L^2\over M_R^2}
+\hat{X}_t^4\left(5\ln{M_R^2\over m_t^2}-29\ln{M_L^2\over m_t^2}\right)
-8\hat{X}_t^5\ln{M_L^2\over M_R^2}\right\}\nn\\
&&+{3\alpha_t m_t^4\over 32\pi^3v^2}\left\{-(2+s_\beta^2)\ln{M_L^2\over
M_R^2}-12s_\beta^2\ln{M_L^2\over m_t^2}\right.\nn\\
&&-2\left[3(14+13s_\beta^2)\hat{X}_t^2
+c_\beta^2(24\hat{X}_t\hat{Y}_t+\hat{Y}_t^2)\right]\ln{M_L^2\over
M_R^2}\nn\\
&&+3\hat{X}_t^4\left(30\ln{M_L^2\over
M_R^2}+35s_\beta^2\ln{M_L^2\over m_t^2}
-23s_\beta^2\ln{M_R^2\over m_t^2}
\right)\nn\\
&&\left.+2\hat{X}_t^2\left[-8c_\beta^2(3\hat{X}_t\hat{Y}_t+\hat{Y}_t^2)
+\hat{X}_t^2(6s_\beta^2\hat{X}_t^2+7c_\beta^2\hat{Y}_t^2)\right]
\ln{M_L^2\over M_R^2}\right\}\ ,
\end{eqnarray}
with $\alpha_t=h_t^2/(4\pi)$.
The two-loop non-logarithmic corrections ($\Delta_{2NNLL}\mh^2$) are
computed
in the next 
Appendix.

\section*{Appendix B: Two-loop threshold corrections to the Higgs quartic
coupling}
\setcounter{equation}{0}
\renewcommand{\theequation}{B.\arabic{equation}}

The threshold corrections to the Higgs quartic self-coupling, $\lambda_H$,
appear in the matching of different effective theories at a given energy
scale. They cannot be calculated by renormalization group methods but
rather must be either computed  by direct diagrammatic calculations or
extracted indirectly from the effective potential. The latter is the
simplest method and is the one we follow in this Appendix.

To illustrate the procedure, we consider first the simpler case of a
unique supersymmetric threshold at $M_S$ ($\gg m_t$) which corresponds to
the common mass scale of all SUSY particles, and obtain the threshold
corrections to $\lambda_H$ (both at $M_S$ and $m_t$) at two-loop order. As
always in this paper we keep only radiative corrections which depend on
the top Yukawa coupling, $h_t$, and/or the strong gauge coupling constant,
$g_s$. Doing this we will reproduce results already presented in
\cite{EZ2}. We discuss later on the case of a hierarchical stop spectrum.
For simplicity we take the stop mixing parameters to be real in this Appendix.

\subsection*{B.1 Degenerate stops}

At the supersymmetric scale, $M_S$, we match the MSSM theory (the `effective
theory' valid above $M_S$) to the SM (the effective theory valid below
$M_S$).  In the effective potential formalism we therefore have to match
the MSSM potential [$V_{MSSM}(m_t)$] to the SM potential [$V_{SM}(\bmt)$].
As indicated, the Higgs field dependence of these functions appears always
through the top quark mass, either directly or through the stop masses,
which are now given by $m_{\sto,{\stw}}^2=M_S^2+m_t^2\pm m_t X_t$. The top
quark mass is a scale-dependent parameter: $m_t$ is the MSSM
$\dr'$-running
mass, while $\bmt$ is the SM running mass in $\ms$ scheme.  Both effective
potentials [$V_{MSSM}(m_t)$ and $V_{SM}(\bmt)$] are known as a perturbative
expansion up to two-loops: $V=V^{(0)}+V^{(1)}+V^{(2)}$. The expression for
$V_{MSSM}$ can be found in \cite{zhang,EZ2}. The two-loop SM effective
potential was first computed in \cite{FJJ}. It can also be extracted from
$V_{MSSM}$ by keeping only the contribution of non-supersymmetric
particles, trading $m_t$ by $\bmt$ and adding an extra piece to account for
the difference between $\dr$ and $\ms$ renormalization schemes. This extra
piece reads
\begin{eqnarray}
&&(\olf)^2~\Delta V_{SM}^{(2)}\ =\ -16 g^2_s
\bmt^4\biggl(1-\ln\frac{\bmt^2}{Q^2}\biggr)\ .
\label{DVSM} 
\end{eqnarray}

As we are interested in the quartic Higgs coupling only, we expand
$V_{MSSM}-V_{SM}$ in powers of $m_t/M_S$ (or $\bmt/M_S$) and keep only the
term that goes like the fourth power of the top quark mass. To proceed
with the computation, we next convert $m_t$ to $\bmt$ using
\be
m^2_t(Q)=\bmt^2(Q)\Biggl\{1-{g_s^2\over 6\pi^2}\biggl[
1+\ls-{M_G X_t\over M_S^2}\biggr]+{3h_t^2\over 32\pi^2} \biggl[
(1+c^2_\beta)\biggl({1\over2}-\ls\biggr)-{1\over 2} \biggr]\Biggr\}\ .
\label{mtshift} 
\ee 
This shift is a loop effect and therefore has different impact on the
different contributions to the $m_t^4$-term depending on their loop-order:
first, it does not matter for the contributions coming from $V^{(2)}$
because the correction would be of three-loop order; second, the quartic
couplings in $V_{MSSM}^{(0)}$ are gauge couplings, and so, transforming
MSSM parameters there to SM ones gives a loop correction which involves
gauge couplings and we are neglecting such effects; finally for
$V^{(1)}$-contributions the correction is a two-loop effect of the same
order of the terms coming from $V^{(2)}$ and must be kept. After this shift
is performed, one simply extracts the threshold correction
$\delta_2\lambda$ from the term
quartic in $\bmt$, with logarithms disregarded (they are renormalization
effects) to obtain
\be
\Delta\mh^2=(\delta_2\lambda) v^2\equiv \Delta_S\ .
\ee

However this is not yet the final result quoted in \cite{EZ2}, eqs.~(29-30)
in that paper. In
addition to the high-energy threshold correction at $M_S$ there is a finite
({\it i.e.} non-logarithmic) low-energy correction coming from
$V^{(2)}_{SM}$ through
\be
\Delta_{\rm th}^{(2)} \mh^2 ={4 m^4_t\over v^2}\left({d\over d
m^2_t}\right)^2 V^{(2)}_{SM}\ .
\label{deltatfor}
\ee
[This kind of correction is zero for the QCD part of $V^{(2)}_{SM}$ or for
$V^{(1)}_{SM}$].

The final result is the sum of the high energy threshold correction
$\Delta_S$ plus the low-energy correction $\Delta_t$ from (\ref{deltatfor}):
\be
\Delta_{\rm th}^{(2)} \mh^2=\Delta_S+\Delta_t,
\ee
with
\begin{eqnarray}
\Delta_S &=&
\frac{\alpha_s m_t^4}{\pi^3v^2}\left[
-2\hxt-\hxt^2+\frac{7}{3}\hxt^3
+\frac{1}{12}\hxt^4-\frac{1}{6}\hxt^5
\right]\nn\\
&+&{3\alpha_t m_t^4\over 16\pi^3 v^2}\Biggl\{
9-16\hxt^2+{13\over 2}\hxt^4
-\frac{1}{2}s_\beta^2\hxt^6\\ 
&+&c^2_{\beta}\biggl[60 K - {13\over2} - {5\pi^2\over3}
-(3+16 K)(4\hxt\hyt+\hyt^2)+ (15-24K) \hxt^2\nn\\
&-&{25\over4}\hxt^4 + (1+4K)\hxt^3\hyt+\biggl({14\over3}
+24 K\biggr)\hxt^2\hyt^2
-\biggl({19\over12}+8K\biggr)\hxt^4\hyt^2\biggr]\Biggr\}\ ,\nn
\label{thrdeghigh}
\end{eqnarray}
and, from (\ref{deltatfor}),
\begin{eqnarray}
\Delta_t &=&
-{3\alpha_t m_t^4\over 16\pi^3 v^2}
\biggl(2+{\pi^2\over3}\biggr)s_\beta^2\ .
\label{thrdeglow}
\end{eqnarray}
We have used $\alpha_s=g_s^2/(4\pi)$, $\alpha_t=h_t^2/(4\pi)$,
$\hat{X}_t=X_t/M_S$, $\hat{Y}_t=Y_t/M_S$ and the
constant $K\simeq-0.1953256$. This result now reproduces exactly that
presented in \cite{EZ2}.

\subsection*{B.2 Hierarchical stops}         

Now we first match at $M_S$ the MSSM as high-energy theory to the $SM +
{\str}$ as the intermediate-energy theory (valid below $M_S=M_L$). The
procedure is similar to the one used in the previous case but with an
expansion of $V_{MSSM} (m_t)-V_{SM + {\stw}}(\bmt)$ in powers of $m_t/M_L$,
$\bmt/M_L$ and $M_R/M_L$.  Again we keep only the quartic power of the top
quark mass in this expansion. Next, notice the argument $\bmt$ for the
potential in the intermediate energy regime: this is because the top quark
mass is insensitive to the ${\str}$ threshold and runs the same above and
below it, that is, with SM RGEs.  The shift from $m_t$ to $\bmt$ in the
one-loop potential contribution is now given by
\be
m^2_t(Q)=\bmt^2(Q)\Biggl\{1-{g_s^2\over 6\pi^2}\biggl[
{1\over 4}+\ln{M_L^2\over Q^2}-2{M_G\over M_L}\hxt\biggr]-{3h_t^2\over
64\pi^2} 
(1+c^2_\beta)\biggl[2\ln{M_L^2\over Q^2}-1\biggr]\Biggr\}\ .
\label{mtshifthier} 
\ee   
We also have to convert from $M_R(Q)$ above $M_L$ to $\overline{M}_R(Q)$
below $M_L$:
\begin{eqnarray}
M_R^2(Q)&=&\overline{M}_R^2(Q)+{g_s^2\over 12\pi^2}\left[
4M_L^2\left(\ln{M_L^2\over Q^2}-1 \right)
+M_R^2\left(1-2\ln{M_L^2\over Q^2}\right)\right]\nn\\
&&
+{h_t^2\over16\pi^2}\left[2s_\beta^2(M_L^2-X_t^2)\left(\ln{M_L^2\over 
Q^2}-1\right)-2c_\beta^2Y_t^2\ln{M_L^2\over Q^2}\right.\nn\\
&&\left.+M_R^2\left(1+\hat{X}_t^2s_\beta^2+{1\over 3}\hat{Y}_t^2c_\beta^2-
2\ln{M_L^2\over Q^2}
\right)
\right]\ ,
\end{eqnarray}
and from $v(Q)$ above $M_L$ to $\overline{v}(Q)$ below $M_L$:
\be
v(Q)=\overline{v}(Q)\left[1-{3g_t^2\over 64\pi^2}\left(\hxt^2-2\ln{m_t^2\over 
Q^2}
\right)
\right]\ .
\ee
For simplicity, in the rest of the formulas we remove the line from
$\overline{M}_R$ in the understanding that it is the parameter of the
intermediate-energy theory.
After throwing away logarithmic terms, one is left with
the following threshold contribution (at the scale $M_S$) to the Higgs
mass: 
\begin{eqnarray} 
\Delta_S &=&\frac{\alpha_s m_t^4}{\pi^3v^2}\left[-{1\over 2}+
{\pi^2\over 3} + \biggl({4\pi^2\over 3}-8\biggr)\hxt-
{3\over 2}\hxt^2
+\biggl({2\pi^2\over 3}-16\biggr)\hxt^3
+\biggl({\pi^2\over 6}+{5\over 4}\biggr)\hxt^4\right.\nn\\
&+&4\hxt^5 \left]
+{3\alpha_t m_t^4\over 16\pi^3
v^2}\right.\Biggl\{ {13\over 2}+{\pi^2\over 3}+
(6+\pi^2)\hxt^2+\biggl({4\pi^2\over 3}-95\biggr)\hxt^4
+(2+\pi^2)\hxt^6s_\beta^2\nn\\ &+&c^2_{\beta}\biggl[
{\pi^2\over 3}-{5\over 2}
+36 K -\biggl(8+{\pi^2\over 3}\biggr)\hxt^2 
+\biggl(8+{4\pi^2\over 3}+72 K\biggr)\hxt\hyt \nn\\
&+&{1\over 9}\hyt^2+(62-\pi^2)\hxt^4
-8(7+27 K)\hxt^3\hyt \nn\\
&+&2\biggl({44\over 9}-{\pi^2\over 3}+18 K\biggr)\hxt^2\hyt^2-
2\biggl({269\over 18}+{\pi^2\over 3}+108 K\biggr)\hxt^4\hyt^2
\biggr]\Biggr\}\ . 
\label{thrhierhigh} 
\end{eqnarray}

For the matching at the intermediate scale $M_R$ we proceed in the same way
with the expansion of $V_{SM + {\str}}(\bmt)-V_{SM}(\bmt)$. 
This expansion gives  a threshold correction to $\mh^2$:
\begin{eqnarray}
\Delta_R &=&
-\frac{\alpha_s m_t^4}{2\pi^3v^2}
(1-\hxt^2)^2+
{3\alpha_t m_t^4\over 8\pi^3 v^2}
(1-\hxt^2)^3s_\beta^2\ .
\label{thrhierint}
\end{eqnarray}

Finally, there is a low-energy correction identical to the one written in
(\ref{thrdeglow}).
The end result  for the two-loop finite correction to the Higgs mass is the
sum of the three contributions just described:
\be
\Delta_{\rm th}^{(2)}
\mh^2\equiv\Delta_{2NNLL}\mh^2=\Delta_S+\Delta_R+\Delta_t\ .
\ee

\section*{Appendix C: On-shell parameters}
\setcounter{equation}{0}
\renewcommand{\theequation}{C.\arabic{equation}}

We write down in the following the useful relations between the
scale-dependent parameters of the top-stop sector and the corresponding
physical on-shell quantities, particularizing general known results to our
case. For simplicity we take the stop mixing parameters to be real in this
Appendix.

The tree-level mass matrix for stops is
\begin{equation}
{\bf M}^2_{\tilt}\
\equiv
\left[\begin{array}{cc}
M_{LL}^2(Q) & M_{LR}^2(Q)\\[2mm]
M_{RL}^2(Q) & M_{RR}^2(Q)
\end{array}\right]
 \simeq\
\left[\begin{array}{cc}
M_L^2+m_t^2  &  m_tX_t\\[2mm]
m_t X_t      &  M_R^2+m_t^2
\end{array}\right]\ ,
\label{stopmat}
\end{equation}  
where we have neglected gauge couplings, which contribute to this matrix
through $D$-terms, and indicated the implicit scale dependence of the
matrix entries.  The one-loop self-energy corrections for top and stops
in the MSSM can be found in \cite{PBMZ,Donini}. In the definition of
physical
parameters, care should be given to the choice of external momentum 
in these self-energies: the scale independence of the quantity in question
crucially depends on that choice. For example, consider the
momentum-dependent loop correction to the stop mass matrix
(\ref{stopmat}):
\begin{equation}
\Delta{\bf M}^2_{\tilt}\ \simeq\
-\left[\begin{array}{cc}
\Pi_{LL}(p^2)  & \Pi_{LR}(p^2) \\[2mm]
\Pi_{RL}(p^2)  & \Pi_{RR}(p^2)
\end{array}\right]\ ,
\label{dstopmat}
\end{equation}  
which also depends on the scale explicitly. Calling 
${\widetilde {{\bf M}}}^2_{\tilt}$ the radiatively corrected stop mass
matrix, ${\bf M}^2_{\tilt}+\Delta{\bf M}^2_{\tilt}$, we find the
following
scale dependence (at one loop):
\begin{eqnarray}
16\pi^2{d\over d\ln Q^2}{\mathrm Tr}\ {\widetilde {{\bf M}}}^2_{\tilt}&=&
h_t^2(M_{LL}^2+2 M_{RR}^2-2p^2)\ ,\nn\\
16\pi^2{d\over d\ln Q^2}{\mathrm Det}\ {\widetilde {{\bf
M}}}^2_{\tilt}&=&
h_t^2\left[3\ {\mathrm Det}\ {\bf M}^2_{\tilt}-p^2(M_{RR}^2+2
M_{LL}^2)\right]\ .
\end{eqnarray}
Using these equations, it is straigthforward to show that the 
momentum-dependent eigenvalues of ${\widetilde {{\bf M}}}^2_{\tilt}$,
being
solutions of the equation
\be
x^2-x\ {\mathrm Tr}\ {\widetilde {{\bf M}}}^2_{\tilt}+{\mathrm
Det}\ {\widetilde
{{\bf
M}}}^2_{\tilt}=0\ ,
\ee
are scale invariant if the external momentum squared is equal to the
eigenvalue itself ({\it i.e.} on-shell).

A stop mixing-angle ($\theta_t$) which includes one-loop radiative corrections
(${\tilde \theta}_t$) can be
defined as the angle of the basis rotation which diagonalizes  
${\widetilde {{\bf M}}}^2_{\tilt}+\Delta{\widetilde {{\bf
M}}}^2_{\tilt}$. As such, it depends on the value of the 
external momentum. The choice of that external momentum required to 
obtain a scale independent definition of the radiatively corrected stop
mixing angle,
${\tilde \theta}_t$, is univocally fixed by demanding 
\be
16\pi^2{d\over d\ln Q^2}\tan
2{\tilde \theta}_t=h_t^2
{(M_{LL}^2+M_{RR}^2-2p^2)\over 2(M_{LL}^2-M_{RR}^2)}\tan 2\theta_t=0\ ,
\ee
which is satisfied for $p^2=(m_{\sto}^2+m_{\stw}^2)/2$.
Once a  scale-independent mixing angle ${\tilde \theta}_t$ has been
obtained in this way, we can
define a physically meaningful, `on-shell' mixing $X_t^{OS}$, by the
relation
\be
\sin 2 {\tilde \theta}_t\equiv {2M_tX_t^{OS} \over M_{\sto}^2-M_{\stw}^2}\
.
\ee

Following these prescriptions, the final results for the physical
parameters in the stop sector are as follows.
For the on-shell stop masses we get
\begin{eqnarray}
M_{\sto}^2&=&m_{\sto}^2(Q)+{g_s^2\over
3\pi^2}\left(2-\ln{M_L^2\over Q^2}\right)M_S^2\nn\\
&&+{h_t^2\over 16\pi^2}\left\{
\left(1-\ln{M_L^2\over Q^2}\right)s_\beta^2M_S^2
+\left(\ln{M_L^2\over
Q^2}-2-{M_R^2\over 
M_S^2}\right)(X_t^2s_\beta^2+Y_t^2c_\beta^2)\right.\nn\\
&&+\pi
c_\beta^2Y_t^2{M_R\over
M_S}+\left.\left[\ln{M_R^2\over
Q^2}-1-\left(\hat{X}_t^2s_\beta^2
+{1\over 2}\hat{Y}_t^2c_\beta^2\right)
\ln{M_L^2\over M_R^2}\right]M_R^2\right\}\ ,\\
&&\nn\\
M_{\stw}^2&=&m_{\stw}^2(Q)+{g_s^2\over
12\pi^2}\left[4\left(1-\ln{M_L^2\over Q^2}\right)M_S^2+ 
\left(5+2\ln{M_L^2\over M_R^2}\right)M_R^2\right]\nn\\
&&+\left.{h_t^2\over
16\pi^2}\right\{
2Y_t^2c_\beta^2\ln{M_L^2\over Q^2}
+2s_\beta^2\left(1-\ln{M_L^2\over Q^2}\right)(M_S^2-X_t^2)\nn\\
&&-\left.\left[1-2\ln{M_L^2\over Q^2}+\hat{X}_t^2s_\beta^2+{1\over
3}\hat{Y}_t^2c_\beta^2
\right]M_R^2\right\}\ ,
\end{eqnarray}
while for the physical stop mixing parameter we find
\begin{eqnarray}
X_t^{OS}&=&X_t(Q)+\left.{g_s^2\over 6\pi^2}\right\{
X_t\left(3\ln{m_t^2\over Q^2}-2\ln{M_L^2\over
Q^2}-{5\over 4}+2\hat{X}_t\right)+2M_S\left(\ln{2M_L^2\over
Q^2}-2\right)\nn\\
&&+\left.X_t\left[-{3\over
2}-9\ln{M_L^2\over
M_R^2}+2\hat{X}_t\left(1-\ln{M_L^2\over M_R^2}\right)-8M_S(2\ln 2-1)
\right]{M_R^2\over 4M_S^2}\right\}\nn\\
&&+\left.{h_t^2\over
16\pi^2}\right\{X_t\left[-{21\over
4}-{3\over2}\ln{m_t^2\over Q^2}+3(1+s_\beta^2)\ln{M_L^2\over
Q^2}+{1\over 2}\ln 2\right.\nn\\
&&\left.+c_\beta^2\left({11\over
4}-{3\over 2}\ln{M_L^2\over m_t^2}-\ln2
\right)+\hat{X}_t^2s_\beta^2(3\ln2-2)+\hat{Y}_t^2c_\beta^2(7S-4-\ln2)
\frac{}{}\right]\nn\\
&&+Y_tc_\beta^2\left[\left(-6+3\ln{M_L^2\over Q^2}+7S+\ln2\right)
+\left(5+2\ln{M_L^2\over M_R^2}-4S-8\ln2\right)
{M_R^2\over M_S^2}\right]\nn\\
&&+\pi X_t\hat{Y}_t^2c_\beta^2{M_R\over M_S}\nn
+X_t\left[-\ln2+c_\beta^2\left(1-2\ln{M_L^2\over 
M_R^2}-2\ln2\right)+\hat{X}_t^2\ln{M_L^2\over M_R^2}
\right.\nn\\
&&+\hat{X}_t^2s_\beta^2(3\ln2-1)\frac{}{}
+\left.\left.\hat{Y}_t^2c_\beta^2\left(-{5\over
2}\ln{M_L^2\over M_R^2}-{17\over
3}+3S+7\ln 2
\right)\right]{M_R^2\over M_S^2}\right\}\ ,
\end{eqnarray}
with
\be
S\equiv{4\over\sqrt{7}}\arctan{1\over\sqrt{7}}.
\ee
Finally, the on-shell top-quark mass is
\begin{eqnarray}
M_t^2&=&m_t^2(Q)\left\{1+
{h_t^2\over
64\pi^2}\left[-19+13c_\beta^2+6\left(\ln{m_t^2\over Q^2}+\ln{M_L^2\over
Q^2}+c_\beta^2\ln{M_L^2\over m_t^2}\right)\right.\right.\nn\\
&&\left.-2\left(1-2\ln{M_L^2\over M_R^2}\right){M_R^2\over 
M_S^2}\right]+{g_s^2\over12\pi^2}\left[{17\over
2}-6\ln{m_t^2\over Q^2}+2\ln{M_L^2\over
Q^2}-4\hat{X}_t\right.\nn\\
&&\left.\left.-\left(1+4\hat{X}_t-
2(1+2\hat{X}_t)\ln{M_L^2\over M_R^2}
\right){M_R^2\over4M_S^2}\right] 
\right\}\ .
\end{eqnarray}

We also need to relate the running vev, $v(Q)$, to some observable, like
the mass of the $W^\pm$. With the type of SUSY spectrum we
consider in this paper, we find
\begin{eqnarray}
\label{v2Q}
v^2(Q) 
&=& {4\over g^2} [M_W^2+{\rm Re}~\Pi_{WW}^T(M_W^2)]\\
&=& 
{4 M^2_W\over g^2}
\left\{1-{3h_t^2s_\beta^2\over32\pi^2}\left[-2\lt+1+\hxt^2
+\hat{X_t}^2\left(3+2\ln{M_R^2\over M_S^2}\right){M_R^2\over M_S^2}
\right]\right\}\ .\nn
\end{eqnarray}
This vev corresponds to the minimum of the one-loop effective potential.
With this definition (others are possible, see \cite{v}) there are no
explicit tadpole contributions in (\ref{v2Q}).
A similar relation, but for the SM vev, $\overline{v}$, reads
\begin{eqnarray}
\overline{v}^2(Q) 
&=& {4 M^2_W\over g^2}
\left[1-{3h_t^2s_\beta^2\over32\pi^2}\left(1-2\lt\right)\right]\ .
\end{eqnarray}
 
In addition, we must consider the relation between the Higgs pole mass,
and the mass obtained from the effective potential. They are related by the
shift
\be
\biggl[-\Pi_{hh}(\mh^2)+\Pi_{hh}(0)\biggl]
={3h_t^2s_\beta^2\over 32\pi^2}\mh^2\left[2\lt+{4\over 3}-\hxt^2
-\hat{X_t}^2\left(3+2\ln{M_R^2\over M_S^2}\right){M_R^2\over M_S^2}
\right]\ .
\ee
Note the partial cancellation that occurs when this correction is
considered together with (\ref{v2Q}).

Some of the threshold corrections and RGEs we present in the main text
can be obtained from the general expressions for self-energies in
\cite{PBMZ} and RGEs in \cite{lahanas} and decoupling SUSY particles in
them. (This works directly for two-point Green functions and through
low-energy theorems for $n$-point ones). We have checked that, whenever
applicable, our results agree with such alternative procedures.

\section*{Acknowledgements}
We thank Andrea Donini, Howie Haber, Andre Hoang and Ren-Jie Zhang for
useful correspondence, Mariano Quir\'os for a careful reading of the
manuscript and Marcos Seco for help with the drawing of Feynman diagrams.



\begin{thebibliography}{99}
%
\bibitem{SUSY}
H.~P.~Nilles, 
Phys.\ Rept.\ {\bf 110} (1984) 1;\\
H.~E.~Haber and G.~L.~Kane, 
Phys.\ Rept.\ {\bf 117} (1985) 75.
%
\bibitem{bound}
P.~Langacker and H.~A.~Weldon,
Phys.\ Rev.\ Lett.\ {\bf 52} (1984) 1377;
\\
H.~A.~Weldon,
Phys.\ Lett.\ B {\bf 146} (1984) 59;
\\
D.~Comelli and J.~R.~Espinosa,
Phys.\ Lett.\ B {\bf 388} (1996) 793
[hep-ph/9607400].
%
\bibitem{higgs1}
S.~P.~Li and M.~Sher,
Phys.\ Lett.\ B {\bf 140} (1984) 339;\\
J.~Ellis, G.~Ridolfi and F.~Zwirner, 
Phys.\ Lett.\ B {\bf 257} (1991) 83;\\
Phys.\ Lett.\ B {\bf 262} (1991) 477;\\
Y.~Okada, M.~Yamaguchi, and T.~Yanagida, 
Prog.\ Theor.\ Phys.\ {\bf 85} (1991) 1;\\
D.~M.~Pierce, A.~Papadopoulos and S.~B.~Johnson, 
Phys.\ Rev.\ Lett.\ {\bf 68} (1992) 3678;\\
M.~Drees and M.~M.~Nojiri, 
Phys.\ Rev.\ D {\bf 45} (1992) 2482;\\
A.~V.~Gladyshev, D.~I.~Kazakov, W.~de Boer, G.~Burkart and R.~Ehret,
Nucl.\ Phys.\ B {\bf 498} (1997) 3
[hep-ph/9603346].
%
\bibitem{radiag} 
M.~S.~Berger, 
Phys.\ Rev.\ D {\bf 41} (1990) 225;\\
H.~E.~Haber and R.~Hempfling, 
Phys.\ Rev.\ Lett.\ {\bf 66} (1991) 1815;\\
M.~A.~D\'{\i}az and H.~E.~Haber, 
Phys.\ Rev.\ D {\bf 46} (1992) 3086;\\
A.~Brignole, 
Phys.\ Lett.\ B {\bf 281} (1992) 284.
%
\bibitem{one}
P.~H.~Chankowski, S.~Pokorski and J.~Rosiek, 
Phys.\ Lett.\ B {\bf 274} (1992) 191;
Nucl.\ Phys.\ B {\bf 423} (1994) 437
[hep-ph/9303309];\\
A.~Yamada,
Phys.\ Lett.\ B {\bf 263} (1991) 233;
Z.\ Phys.\ C {\bf 61} (1994) 247;\\
A.~Dabelstein, 
Z.\ Phys.\ C {\bf 67} (1995) 495 [hep-ph/9409375].
%
\bibitem{PBMZ} 
D.~M.~Pierce, J.~A.~Bagger, K.~Matchev and R.~Zhang,
Nucl.\ Phys.\  {\bf B491} (1997) 3
[hep-ph/9606211].
%
\bibitem{radRG} 
R.~Barbieri, M.~Frigeni and M.~Caravaglios, 
Phys.\ Lett.\ B {\bf 258} (1991) 167;\\
Y.~Okada, M.~Yamaguchi, and T.~Yanagida, 
Phys.\ Lett.\ B {\bf 262} (1991) 54;\\
J.~R.~Espinosa and M.~Quir\'{o}s,
Phys.\ Lett.\ B {\bf 266} (1991) 389;\\
K.~Sasaki, M.~Carena and C.~E.~M. Wagner, \NPB{381}{92}{66};\\
H.~E.~Haber and R.~Hempfling, 
Phys.\ Rev.\ D {\bf 48} (1993) 4280
[hep-ph/9307201].
%
\bibitem{CEQR} J.~A.~Casas, J.~R.~Espinosa, M.~Quir\'os and A.~Riotto, 
Nucl.\ Phys.\ B {\bf 436} (1995) 3
[hep-ph/9407389].
%
\bibitem{H2} R.~Hempfling and A.~H.~Hoang, 
Phys.\ Lett.\ B {\bf 331} (1994) 99
[hep-ph/9401219].
%
\bibitem{CEQW} M.~Carena, J.~R.~Espinosa, M.~Quir\'os and C.~E.~M.~Wagner,
Phys.\ Lett.\ B {\bf 355} (1995) 209
[hep-ph/9504316].
%
\bibitem{CQW} M.~Carena, M.~Quir\'os and C.~E.~M.~Wagner,
Nucl.\ Phys.\ B {\bf 461} (1996) 407
[hep-ph/9508343].
%
\bibitem{H3} H.~E.~Haber, R.~Hempfling and A.~H.~Hoang, 
Z.\ Phys.\ C {\bf 75} (1997) 539
[hep-ph/9609331].
%
\bibitem{zhang}
R.-J.~Zhang,
Phys.\ Lett.\  {\bf B447} (1999) 89
[hep-ph/9808299].
%
\bibitem{EZ}
J.~R.~Espinosa and R.-J.~Zhang,
JHEP {\bf 0003} (2000) 026
[hep-ph/9912236].
%
\bibitem{EZ2}
J.~R.~Espinosa and R.-J.~Zhang,
Nucl.\ Phys.\  {\bf B586} (2000) 3
[hep-ph/0003246].
%
\bibitem{hollik} 
S.~Heinemeyer, W.~Hollik and G.~Weiglein, 
Phys.\ Rev.\ D {\bf 58} (1998) 091701
[hep-ph/9803277];
Eur.\ Phys.\ J.\ C {\bf 9} (1999) 343
[hep-ph/9812472];
Phys.\ Lett.\ B {\bf 440} (1998) 296
[hep-ph/9807423].
%
\bibitem{CH3W2} 
M.~Carena, H.~E.~Haber, S.~Heinemeyer, W.~Hollik, C.~E.~Wagner and
G.~Weiglein,
Nucl.\ Phys.\  {\bf B580} (2000) 29
[hep-ph/0001002].
%
\bibitem{Efft}
H.~Georgi,
Ann.\ Rev.\ Nucl.\ Part.\ Sci.\  {\bf 43} (1993) 209;
\\
A.~G.~Cohen,
{\it Prepared for Theoretical Advanced Study Institute (TASI
93) in Elementary Particle Physics: The Building Blocks of
Creation - From
Microfermius to Megaparsecs, Boulder, CO, 6 Jun - 2 Jul 1993};
\\
D.~B.~Kaplan,
[nucl-th/9506035];
\\
A.~Pich,
[hep-ph/9806303].
%
%
\bibitem{nyffeler}
A.~Nyffeler and A.~Schenk,
Annals Phys.\ {\bf 241} (1995) 301
[hep-ph/9409436].
%
\bibitem{MSSMdec}
A.~Dobado, M.~J.~Herrero and S.~Pe\~naranda,
Eur.\ Phys.\ J.\ C {\bf 7} (1999) 313
[hep-ph/9710313];
Eur.\ Phys.\ J.\ C {\bf 12} (2000) 673
[hep-ph/9903211];
Eur.\ Phys.\ J.\ C {\bf 17} (2000) 487
[hep-ph/0002134];\\
H.~E.~Haber, M.~J.~Herrero, H.~E.~Logan, S.~Penaranda, S.~Rigolin and
D.~Temes,
Phys.\ Rev.\ D {\bf 63} (2001) 055004
[hep-ph/0007006].
%
\bibitem{martin}
S.~P.~Martin and M.~T.~Vaughn,
Phys.\ Lett.\ B {\bf 318} (1993) 331
[hep-ph/9308222].
%
\bibitem{DR} W.~Siegel, 
Phys.\ Lett.\ B {\bf 84} (1979) 193;\\
D.~M.~Capper, D.~R.~T.~Jones and P.~van~Nieuwenhuizen, 
Nucl.\ Phys.\ B {\bf 167} (1980) 479;\\
I.~Jack, D.~R.~T.~Jones, S.~P.~Martin, M.~T.~Vaughn and
Y.~Yamada, 
Phys.\ Rev.\ D {\bf 50} (1994) 5481
[hep-ph/9407291].
%
\bibitem{aoki}
K.~I.~Aoki, Z.~Hioki, M.~Konuma, R.~Kawabe and T.~Muta,
Prog.\ Theor.\ Phys.\ Suppl.\ {\bf 73} (1982) 1.
%
\bibitem{RGE2L} 
M.~E.~Machacek and M.~T.~Vaughn,
Nucl.\ Phys.\  {\bf B249} (1985) 70.
%
%
\bibitem{SZ}
A.~Sirlin and R.~Zucchini, 
Nucl.\ Phys.\ B {\bf 266} (1986) 389.
%
\bibitem{multiV}
M.~B.~Einhorn and D.~R.~Jones,
Nucl.\ Phys.\ B {\bf 230} (1984) 261;\\
C.~Ford and C.~Wiesendanger,
Phys.\ Rev.\ D {\bf 55} (1997) 2202
[hep-ph/9604392];
Phys.\ Lett.\ B {\bf 398} (1997) 342
[hep-th/9612193];\\
M.~Bando, T.~Kugo, N.~Maekawa and H.~Nakano,
Prog.\ Theor.\ Phys.\ {\bf 90} (1993) 405
[hep-ph/9210229];\\
H.~Nakano and Y.~Yoshida,
Phys.\ Rev.\ D {\bf 49} (1994) 5393
[hep-ph/9309215].
%
\bibitem{multiV1}
J.~A.~Casas, V.~Di Clemente and M.~Quiros,
Nucl.\ Phys.\ B {\bf 553} (1999) 511
[hep-ph/9809275].
%
\bibitem{FJJ}
C.~Ford, I.~Jack and D.~R.~Jones,
Nucl.\ Phys.\  {\bf B387} (1992) 373; Erratum-ibid. {\bf B504}
(1997) 551.
%
\bibitem{Donini}
A.~Donini,
Nucl.\ Phys.\  {\bf B467} (1996) 3
[hep-ph/9511289].
%
%
\bibitem{v}
R.~Hempfling and B.~A.~Kniehl,
Phys.\ Rev.\ {\bf D 51} (1995) 1386
[hep-ph/9408313].
%
\bibitem{lahanas}
A.~B.~Lahanas and K.~Tamvakis,
Phys.\ Lett.\  {\bf B348} (1995) 451
[hep-ph/9412281].
%
\end{thebibliography}
\end{document}